\documentclass[acmsmall,review=True]{acmart}

\usepackage{booktabs} % For formal tables
\usepackage{amsmath}
\usepackage{algorithm}
\usepackage{multirow}
\usepackage{flushend}
\usepackage{enumitem}
\usepackage[T1]{fontenc}
\usepackage[utf8]{inputenc}
\usepackage{algorithmic}
\usepackage{xcolor}

\newtheorem{example}{Example}[section]
\newtheorem*{proof*}{Proof}
\newtheorem*{claim*}{}

\AtBeginDocument{%
  \providecommand\BibTeX{{%
    \normalfont B\kern-0.5em{\scshape i\kern-0.25em b}\kern-0.8em\TeX}}}

\setcopyright{acmcopyright}
\copyrightyear{2021}
\acmYear{2021}
\acmDOI{10.1145/3366423.3380294}

\acmJournal{TIST}
\acmVolume{37}
\acmNumber{4}
\acmArticle{111}
\acmMonth{8}

\begin{document}
\title{CDSM: \underline{C}ascaded \underline{D}eep \underline{S}emantic \underline{M}atching on Textual Graphs Leveraging Ad-hoc Neighbor Selection}
 
\author{Jing Yao}
\email{jingyao@microsoft.com}
\affiliation{%
  %\institution{School of Information, Renmin University of China}
  \institution{Microsoft Research Asia}
  \city{Beijing}
  \postcode{100080}
}
\author{Zheng Liu}
\authornote{Corresponding author.}
\email{zhengliu@microsoft.com}
\affiliation{%
  \institution{Microsoft Research Asia}
  \city{Beijing}
  \postcode{100080}
}
\author{Junhan Yang}
\authornote{Work was done during Junhan Yang's internship in Microsoft.}
\email{t-junhanyang@microsoft.com}
\affiliation{%
  \institution{University of Science and Technology of China}
  \city{Hefei}
  \postcode{230026}
}
\author{Zhicheng Dou}
\email{dou@ruc.edu.cn}
\affiliation{%
  \institution{Gaoling School of Artificial Intelligence, Renmin University of China}
  \city{Beijing}
  \postcode{100872}
}
\author{Xing Xie}
\email{xingx@microsoft.com}
\affiliation{%
  \institution{Microsoft Research Asia}
  \city{Beijing}
  \postcode{100080}
}
\author{Ji-Rong Wen}
\email{jrwen@ruc.edu.cn}
\affiliation{%
  \institution{Engineering Research Center of Next-Generation Intelligent Search and Recommendation, Ministry of Education}
  \city{Beijing}
  \postcode{100872}
}

\fancyhead{}
\renewcommand{\shortauthors}{Jing Yao, Zheng Liu, Junhan Yang and Zhicheng Dou} 
 
\begin{abstract}
Deep semantic matching aims to discriminate the relationship between documents based on deep neural networks. In recent years, it becomes increasingly popular to organize documents with a graph structure, then leverage both the intrinsic document features and the extrinsic neighbor features to derive discrimination. Most of the existing works mainly care about how to utilize the presented neighbors, whereas limited effort is made to filter appropriate neighbors. We argue that the neighbor features could be highly noisy and partially useful. Thus, a lack of effective neighbor selection will not only incur a great deal of unnecessary computation cost, but also restrict the matching accuracy severely.

In this work, we propose a novel framework, \textbf{C}ascaded \textbf{D}eep \textbf{S}emantic \textbf{M}atching (\textbf{CDSM}), for accurate and efficient semantic matching on textual graphs. CDSM is highlighted for its two-stage workflow. In the first stage, a lightweight CNN-based ad-hod neighbor selector is deployed to filter useful neighbors for the matching task with a small computation cost. We design both one-step and multi-step selection methods. In the second stage, a high-capacity graph-based matching network is employed to compute fine-grained relevance scores based on the well-selected neighbors. It is worth noting that CDSM is a generic framework which accommodates most of the mainstream graph-based semantic matching networks. The major challenge is how the selector can learn to discriminate the neighbors' usefulness which has no explicit labels. To cope with this problem, we design a weak-supervision strategy for optimization, where we train the graph-based matching network at first and then the ad-hoc neighbor selector is learned on top of the annotations from the matching network. We conduct extensive experiments with three large-scale datasets, showing that CDSM notably improves the semantic matching accuracy and efficiency thanks to the selection of high-quality neighbors. The source code is released at https://github.com/jingjyyao/CDSM.
\end{abstract}




\maketitle

\section{Introduction}\label{sec:intro}
Deep semantic matching aims to discriminate the relationship between documents based on deep neural networks \cite{huang2013learning,shen2014learning,guo2016deep}. It plays a critical role in today's intelligent web services, such as recommendation systems, search engines, and online advertising systems~\cite{xu2018deep}. Conventional semantic matching models mainly analyze the intrinsic features of the documents. Thanks to the increasing capacity of deep neural networks, a series of advanced document encoders have been proposed, especially those based on pre-trained language models \cite{reimers2019sentence,khattab2020colbert,luan2020sparse}.

\setlength{\textfloatsep}{10pt}
\begin{figure}
    \centering
    \includegraphics[width=0.7\columnwidth]{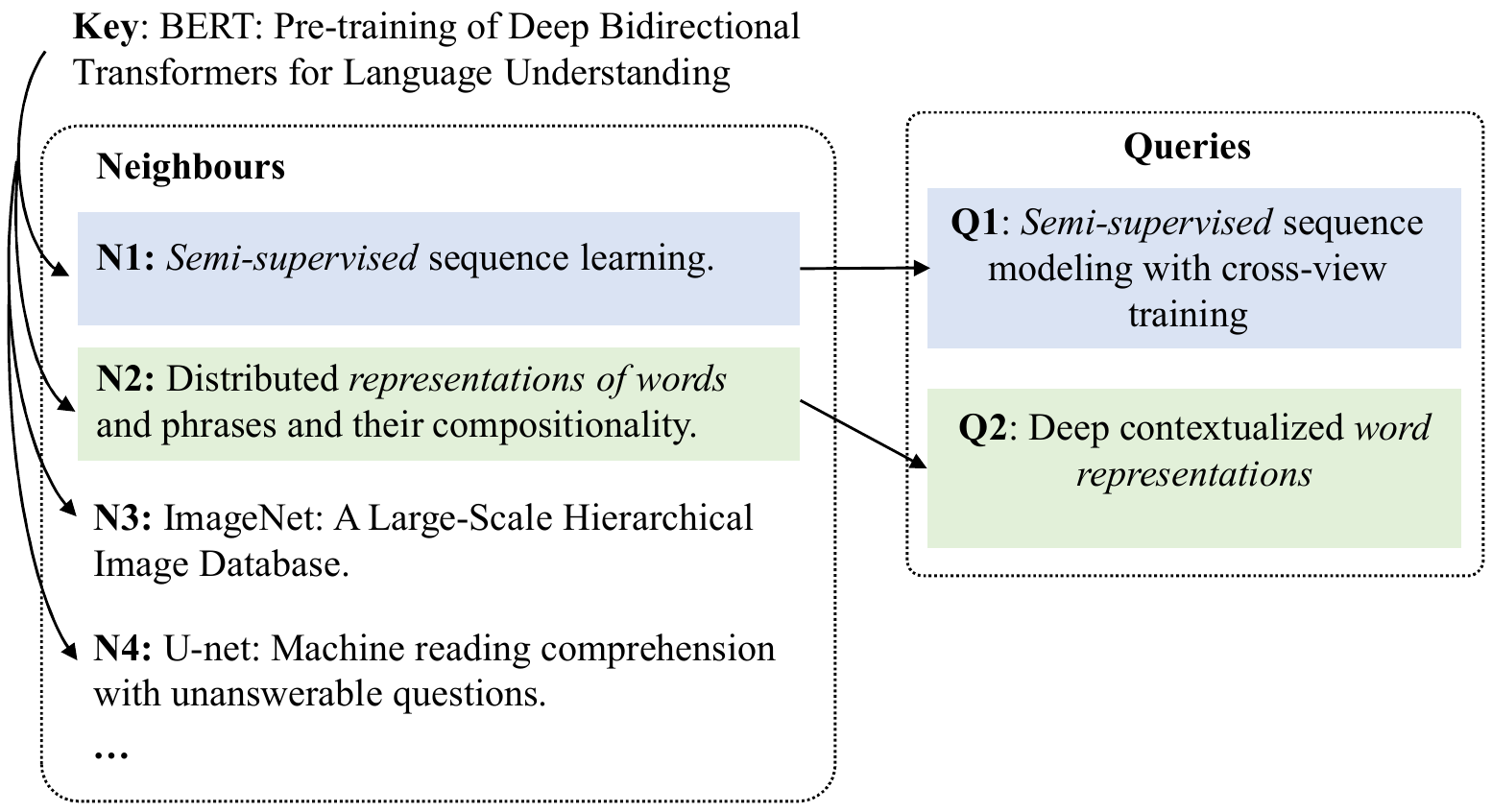}
    \caption{Example of ad-hoc neighbor selection. The key document is linked to multiple neighbors (papers in its reference) of different semantics. While estimating its relationship with Q1, N1 will be selected as both documents are about \textit{semi-supervised learning}; while estimating the relationship with Q2, N2 will be selected due to the correlation with \textit{word representation}.} 
    \label{fig:1}
\end{figure}

\subsection{Semantic Matching on Textual Graphs}
Many textual datasets can be naturally organized with graph structures, e.g., the webpages of online products can be linked based on users' web browsing behaviors, and the academic literature can be linked based on their citation relationships. In recent years, it becomes increasingly popular to discriminate the matching relationship between two documents on such textual graphs \cite{Zhu2021TextGNN,Li2021AdsGNN,zhang2020mira,Yang2021GraphFormer}, where both the intrinsic document features and the extrinsic neighbor features are jointly leveraged. Existing works mainly emphasize the design of graph-based matching models, whereas limited effort is dedicated to the selection of appropriate neighbors. It's usually assumed that most of the neighbors are informative, and semantic matching may always benefit from the incorporation of neighbor features. As a result, existing methods~\cite{Zhu2021TextGNN,Yang2021GraphFormer} either use all the available neighbors if the computation capacity allows, or heuristically sample a subset of neighbors for efficient computation.

We argue that \textbf{the neighbors could be highly noisy and partially useful, thus appropriate neighbor selection is vital for graph-based semantic matching}. Contradicted to the common assumption, we empirically find the following properties. \emph{1) The neighbor features can be highly \textbf{noisy}: in many situations, only a small fraction of neighbors could actually contribute to the current semantic matching}. \emph{2) The neighbor feature's usefulness is \textbf{conditional}: a neighbor may contribute to the semantic matching given one particular target document, but it may become useless when dealing with another target document (as Example~\ref{eq:1}).} As such, a lack of effective neighbor selection will severely restrict the performance of graph-based semantic matching. The irrelevant neighbors will not only {take a huge amount of unnecessary computation cost}, but also {introduce strong background noise which deteriorates the matching accuracy}.

\subsection{Our Work}
In this paper, we propose the \textbf{C}ascaded \textbf{D}eep \textbf{S}emantic \textbf{M}atching (\textbf{CDSM}) framework to address the above challenges. It completes the semantic matching task in two consecutive steps. First, an ad-hoc neighbor selection step is performed to select the optimal subset of neighbors w.r.t. the given documents. By focusing on these selected neighbors which are truly informative, the subsequent semantic matching step can be accomplished with both high accuracy and high efficiency, as background noise and unnecessary computation costs from the irrelevant neighbors can be avoided.

$\bullet$ \textbf{Ad-hoc Neighbor Selection}. In CDSM, the neighbor selection is performed in an ad-hoc manner. It is empirically found that {merely a small number of neighbors may accurately contribute to the semantic matching task between two specific documents; and an informative neighbor in one matching task will probably become useless in another task}\footnote{check Section \ref{sec:problem_definition} for the empirical analysis}. Therefore, static neighbor selection will be inappropriate. Instead of assigning fixed neighbors to each document, CDSM makes \textbf{Ad-hoc} neighbor selection, i.e., neighbors are selected w.r.t the counterpart to be matched. Specifically, given a document and its matching counterpart, CDSM aims to identify the neighbors that are closely related to the counterpart. Only the neighbors relevant to the counterpart will be preserved for the subsequent semantic matching task, while the irrelevant ones will be filtered out.
We use a concrete example to illustrate the underlying intuition. 

\begin{example}\label{eq:1}
As shown in Figure \ref{fig:1}, the key document ``BERT: Pre-training of Deep Bidirectional Transformers for Language Understanding'' should be linked to all documents within its reference list. N1 is about ``semi-supervised learning'', and N2 is about the ``representation of words''; meanwhile, there are many more neighbors about other topics, like ``ImageNet dataset'', ``machine reading comprehension'', etc. To determine whether the query document Q1 has a citation link with the key document, the ad-hoc neighbor selector will choose N1 to contribute useful information, rather than N2. Both N1 and Q1 are about ``semi-supervised learning''. Based on such supporting evidence, the connection between the key document and Q1 can be positively determined with high confidence. However, given another query document Q2, the neighbor N2 becomes a plausible choice. Both Q2 and N2 are about ``word representation'', based on which the connection between the key document and Q2 can be positively determined with high confidence.
\end{example}

The neighbor selector is also designed to be \textbf{lightweight}. Given that each document may be linked to a vast number of neighbors, it will be infeasible to identify their utilities with heavy-loaded functions. In CDSM, the neighbor selector makes use of lightweight estimation networks, whose inference cost will be small. It is also experimentally validated that the highly simplified backbone networks are already sufficiently accurate to identify useful neighbors. As a result, the computation overhead of neighbor selection will be small enough, making CDSM comparably efficient as the existing methods based on heuristic neighbor sampling.

The subsequent semantic matching is performed based on the well-selected neighbors from the first stage. As discussed, both accuracy and efficiency of the semantic matching will benefit from such a selection, thanks to the elimination of background noise and unnecessary computation costs. It is worth noting that CDSM is a \textbf{Generic} framework, where the mainstream graph-based matching models (e.g., the recent works which combine GNNs and pre-trained language models \cite{zhang2020mira,Li2021AdsGNN,Zhu2021TextGNN}) can be seamlessly incorporated as the backbone of the matching network. 

$\bullet$ \textbf{Training via weak-supervision}. One major challenge for CDSM is that there are no explicit measurements of the neighbors' usefulness, which hinders the training of the neighbor selector. In our work, we develop a weak-supervision strategy, where the neighbor selector is trained on top of weak annotations obtained from the semantic matching model. Intuitively, given a pair of documents Q and K, a useful neighbor of Q should provide additional evidence to support the correlation between Q and K. In other words, a useful neighbor will strengthen the correlation between Q and K. Based on such an intuition, we make a formalized definition, which serves as the criterion of whether a neighbor is useful in a specific semantic matching task.
\begin{definition}\label{def:1}
Given a pair of documents $Q$ and $K$, and the function $\mathcal{M}$($\cdot$) which measures the semantic correlation between $Q$ and $K$. A neighbor $Q.N_i$ is useful, if the semantic correlation between $Q$ and $K$ is improved when $Q$ is aggregated with $Q.N_i$: $\mathcal{M}(Q\oplus{Q.N_i},K) > \mathcal{M}(Q,K)$ (``$\oplus$'' is the aggregation operator).
\end{definition}
With the definition of neighbor usefulness, we develop a weak-supervision algorithm for CDSM. First, the graph-based semantic matching network $\mathcal{M}$ is trained with the neighbors sampled by heuristics. Second, the well-trained semantic matching network makes annotation for the usefulness of each neighbor $Q.N_i/K.N_i$. Finally, the neighbor selector is trained based on the annotation results, where it learns to discriminate the highly useful neighbors from the less useful ones.

Extensive experimental studies are conducted with three large-scale datasets: DBLP, Wiki and Bing Ads\footnote{The first two are widely used open benchmark datasets, and the third one is a massive industrial dataset collected by Bing Search.}. The CDSM's effectiveness is verified from two perspectives. Firstly, as a generic framework, it notably improves the accuracy of a variety of graph-based semantic matching models. Secondly, it achieves competitive running efficiency as high-quality neighbors can be effectively selected with a small computation cost.

To summarize, our major contributions are listed as follows:
\begin{itemize}
    \item We identify the importance of neighbor selection in the graph-based semantic matching task.
    \item We propose a novel generic framework CDSM. With the selection of high-quality neighbors, CDSM achieves both high accuracy and high efficiency for graph-based semantic matching.
    \item We design the weak-supervision strategy to effectively train the neighbor selector on top of the semantic matching network's annotations.
    \item We perform comprehensive experimental studies, whose results verify the effectiveness and efficiency of CDSM as a generic framework.
\end{itemize}

The rest of the paper is organized as follows. The related works are reviewed in Section~\ref{sec:related_work}. The graph-based semantic matching problem is formally defined and empirically analyzed in Section~\ref{sec:problem_definition}. The detailed methodology of CDSM is elaborated in Section~\ref{sec:framework}. The experimental studies are discussed in Section~\ref{sec:experimental_setting} and \ref{sec:results}. Finally, the whole work is concluded in Section~\ref{sec:conclusion}.

\section{Related Works}\label{sec:related_work}
Semantic matching between documents is a fundamental problem in information retrieval and recommendation systems. Conventional works mainly rely on the intrinsic document features, e.g., bag-of-words \cite{robertson2009probabilistic}, latent semantic analysis \cite{landauer1998introduction} and topic modeling \cite{blei2003latent}. In recent years, various neural text encoders have been extensively explored for this task. In \cite{shen2014learning,yang2016hierarchical}, convolution neural networks are utilized to capture the local contextual patterns of the input texts; and in \cite{palangi2016deep,Seo2017BiDAF}, recurrent neural networks are employed to encode the sequential relationship between tokens. The latest works are usually built upon the large-scale pre-trained language models, like BERT and RoBERTa \cite{Devlin2019BERT,Liu2019Roberta,reimers2019sentence}. The semantic matching accuracy can be significantly improved thanks to the high-quality deep contextualized text representations.

Actually, many textual datasets can be organized in the form of a graph. For example, in sponsored search, advertisements can be linked together based on users' co-click behaviors \cite{Zhu2021TextGNN,Li2021AdsGNN}. Similarly, online articles and webpages can be connected as a graph according to their mutual linkage relationships. In recent years, it becomes increasingly popular to conduct fine-grained semantic matching on such textual graphs~\cite{Zhu2021TextGNN,Li2021AdsGNN,Yang2021GraphFormer,zhang2020mira,Hu2020RecoGNN,Arora2020KGcompletionSurvey,Wang2021KGCompletionSurvey}, leveraging both the intrinsic document features and the extrinsic neighbor features. The typical approaches \cite{hamilton2017inductive,ying2018graph} will firstly encode each document node into latent representations; and then make use of graph neural networks to aggregate the graph neighbor information. The latest works usually combine graph neural networks with pre-trained language models \cite{Zhu2021TextGNN,Li2021AdsGNN}, where the underlying semantics of each individual document can be captured more effectively. To facilitate the in-depth interaction between the textual features and graph structures, Yang et al.~\cite{Yang2021GraphFormer} propose GraphFormers: the GNN components are nested into each layer of the transformer. In such a way, each document's representation can be contextualized by involving the graph information.

Despite the achieved progress so far, these works mainly focus on designing a better model to aggregate existing neighbors (all neighbors or those selected heuristically), whereas limited effort is made for the selection of truly informative neighbors. Empirically, we have found that 1) the neighbor features can be noisy and 2) the usefulness of a neighbor is conditional, different for different matching counterparts. In this paper, we are committed to solving the problem of effective neighbor selection to improve both the accuracy and efficiency of graph-based semantic matching.

\section{Problem Definition}\label{sec:problem_definition}
This section covers the following issues: 1) the definition of graph-based semantic matching, 2) the empirical analysis of how neighbor selection matters in graph-based semantic matching, and 3) the definition of neighbor selection, which plays a central role in our CDSM framework.

\subsection{Graph-based Semantic Matching}
Semantic matching is a critical issue in information retrieval and natural language processing. Given a query document $Q$ and a key document $K$, a semantic matching model predicts the relevance score between the two documents based on their textual features. Since many real-world data can be organized as graphs, the graph-based semantic matching goes beyond by leveraging the textual features from both the target nodes (i.e. $Q$ and $K$) and their linked neighbors on the graph. In this place, we define the graph-based semantic matching task in the form of a typical ranking problem.

\begin{definition}\label{def:2}
{(Graph-based Semantic Matching)} Given a pair of documents: query $Q$ and key $K$, together with their neighbor sets $Q.N$ and $K.N$, the graph-based semantic matching model $\mathcal{M}$ learns to predict the relevance score between the query and key as:
\begin{equation}
    m(Q,K) = \mathcal{M}(Q\oplus{Q.N}, K\oplus{K.N}),
\end{equation}
(``$\oplus$'' is the aggregation operator), such that a positive key $K^+$ can be ranked higher than a negative key $K^-$: $m(Q,K^+) > m(Q,K^-)$.
\end{definition}

Although the neighbor features may provide complementary information to the semantic matching task, they should be utilized with caution due to the following defects. First, the neighbor features are prone to strong noise: many of the neighbors may contribute little useful information to the semantic matching task. Secondly, the usefulness of a neighbor is conditional: a neighbor can be helpful to the semantic matching task between a query and a specific key document, but turns useless when dealing with other keys. In the following discussion, an empirical analysis is presented to demonstrate the neighbors' impacts.

\setlength{\textfloatsep}{10pt}
\begin{figure}
    \centering
    \includegraphics[width=0.6\columnwidth]{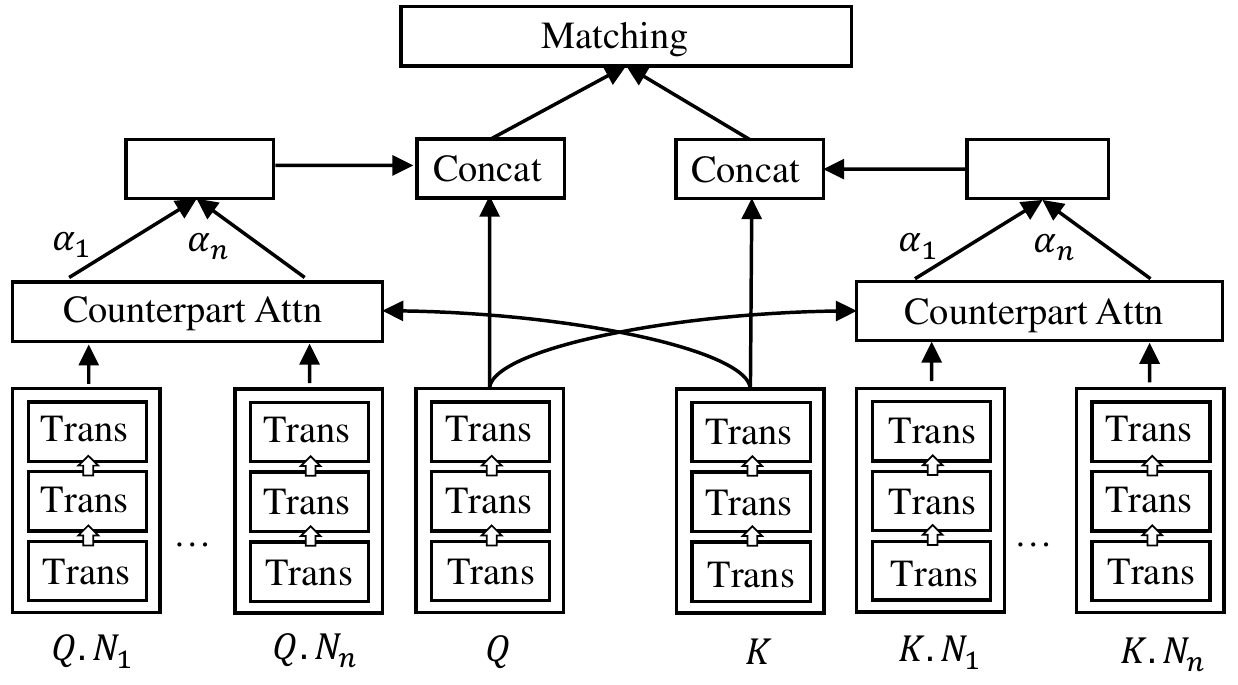}
    \caption{TextGNN with counterpart attention. The center document $Q, K$ and their neighbors are encoded by the pre-trained language model (with stacked transformers) at first; the neighbors are then aggregated based on the attention of the matching counterpart.}
    \label{fig:2}
\end{figure}

\begin{figure}
    \centering
    \includegraphics[width=0.6\linewidth]{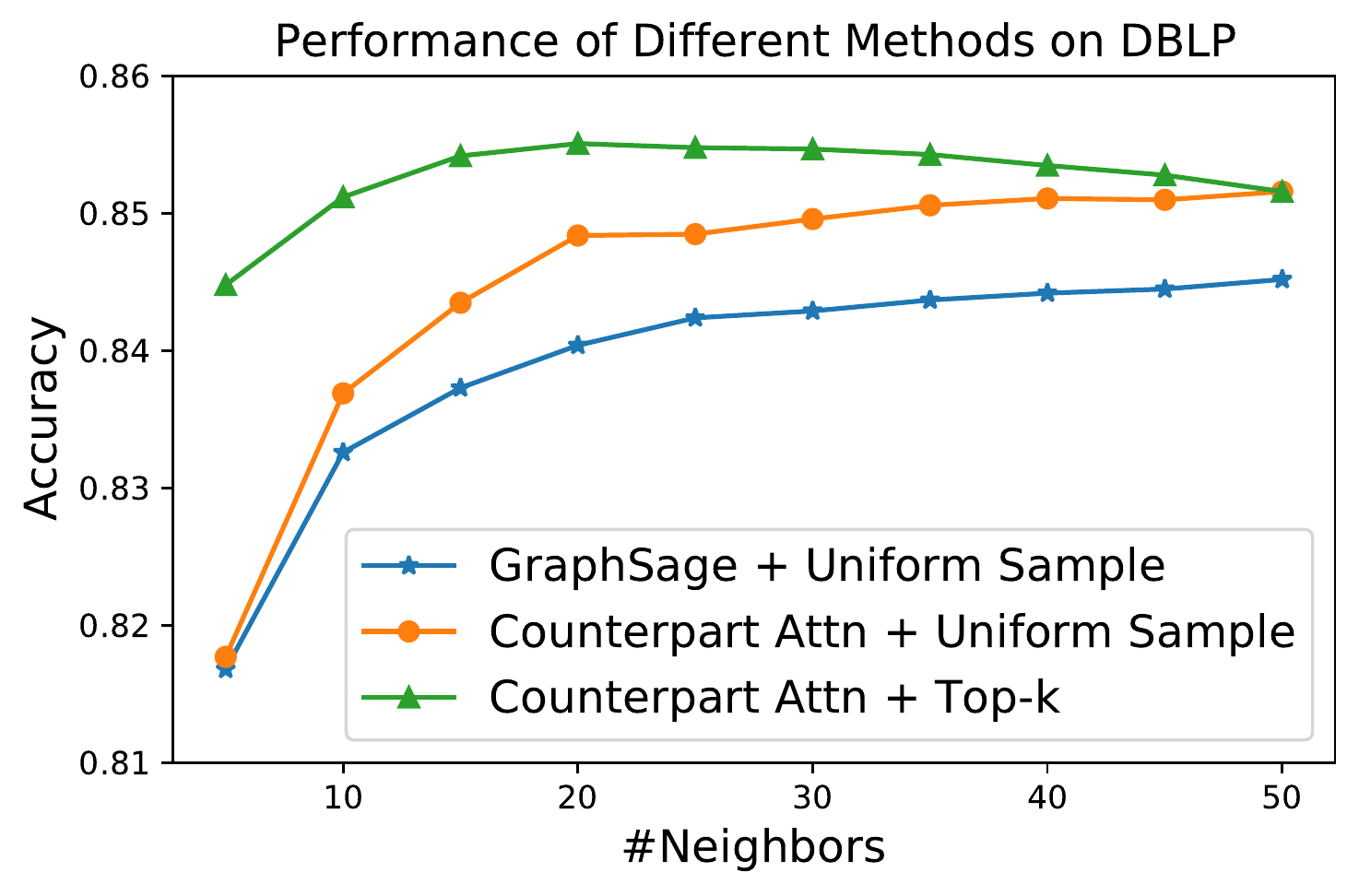}
    \caption{Case analysis on DBLP. The horizontal axis shows the number of neighbors utilized in each semantic matching task, and the vertical axis shows the semantic matching accuracy (measured by precision@1).}
    \label{fig:preliminary_analysis}
\end{figure}

\subsection{Neighbors' Impact: An Empirical Analysis with TextGNN}
We take one of the latest graph-based semantic matching methods, TextGNN \cite{Zhu2021TextGNN}, for our analysis. The TextGNN framework uses the pre-trained language model BERT as the text encoder, and graph neural networks (like GraphSage) as the graph aggregator. The input documents, i.e. the query/key and their neighbors, are encoded by the text encoder at first. Then, the text embeddings are aggregated by the graph aggregator to get the final representation for semantic matching. In our work, we make an analysis based on the following adaptations of TextGNN.

\textbf{First}, we introduce the ``counterpart attention'' as shown in Figure \ref{fig:2}. The ``counterpart'' means the document to be matched: the query's counterpart is the key, while the key's counterpart is the query. To identify the neighbors that truly contribute to the current matching task, the neighbors are scored and aggregated based on their attention with the counterpart document:
\begin{equation}
\text{AGG}(N) = \sum_i \alpha_i * \theta_{N_i}, ~
\alpha_i = \mathrm{CP-ATT}(\theta_{N_i}, \theta_{CP}).
\end{equation}
$\text{AGG}(\cdot)$ is the neighbor aggregation function and $\text{CP-ATT}(\cdot)$ means the counterpart attention. $\theta_{N_i}$ and $\theta_{CP}$ are the embeddings of the neighbor $N_i$ (could be $Q.N_i$ or $K.N_i$) and the counterpart document respectively. With this adaptation, TextGNN becomes more robust to noisy neighbors, as truly useful neighbors for the counterpart can be highlighted with higher attention weights.

\textbf{Second}, we select a subset of neighbors for graph-based semantic matching. A neighbor is selected if it is ranked within the ``Top-$k$'' positions w.r.t. the counterpart attention scores. By this means, the quality of neighbor features can be improved as the potentially irrelevant neighbors can be largely removed, therefore leading to higher semantic matching accuracy.

We perform experimental analysis with the DBLP dataset (refer to Section \ref{sec:experimental_setting} for details). A total of three alternatives are compared: (1) The original TextGNN with GraphSage aggregator (using mean-pooling for implementation), and uniformly sampled neighbors. (2) The adapted TextGNN with counterpart attention, and uniformly sampled neighbors. (3) The adapted TextGNN with counterpart attention, and the Top-$k$ neighbors. All neighbors are sampled from the same candidate set, and each document owns at most 50 neighbors. The analysis results are shown in Figure \ref{fig:preliminary_analysis}. The semantic matching accuracy (measured by Precision@1, with 1 positive key and 29 negative keys) is checked when different numbers of neighbors are incorporated. The following properties can be observed from the shown results.

$\bullet$ {There is indeed strong noise within the neighbor features.} The adapted TextGNN with Top-$k$ neighbors consistently outperforms the one with uniformly sampled neighbors, which indicates that the subset of Top-$k$ neighbors is more useful. Besides, the accuracy of TextGNN with the Top-$k$ neighbors reaches its apex when $k$ is merely around 20; it goes down when more neighbors are introduced. It is probably because the additional neighbors (thereafter the apex) are less relevant, which provides little useful information but noisy features to the semantic matching task.

$\bullet$ {The usefulness of neighbors is conditional, dependent on the given counterpart}. For both methods with the uniformly sampled neighbors, the adapted TextGNN with counterpart attention outperforms the original TextGNN consistently. Such an observation indicates that a neighbor is likely to contribute more to the semantic matching task if it is relevant to the matching counterpart.

Both findings are consistent with our statements about the neighbor features: noisy and conditionally useful. Without an effective neighbor selection mechanism, the graph-based semantic matching could be inaccurate and inefficient due to the introduction of useless neighbors. Notice that although the ``counterpart attention based Top-$k$ neighbor selection'' improves the accuracy, it has low feasibility in practice as the entire neighbors still need to be encoded by heavy-loaded text encoders. A practical selection mechanism will be introduced in our subsequent discussion.

\subsection{Neighbor Selection}
Given the discussed properties of the neighbor features, it is important to select the subset of neighbors which are truly helpful to the specific semantic matching task. In this paper, we aim to learn a neighbor selector to complete the selection defined as follows.

\begin{definition}
(Neighbor Selection) Given the semantic matching task between the query document $Q$ and key document $K$ (whether $K$ is positive or negative is unknown), the neighbor selector $\mathcal{S}(\cdot)$ learns to identify the subsets of neighbors:
\begin{equation}
    Q.\tilde{N} \leftarrow \mathcal{S}(Q.N|Q,K), ~K.\tilde{N} \leftarrow \mathcal{S}(K.N|Q,K);
\end{equation}
such that the semantic matching accuracy can be optimized on top of the selection result: $Q\oplus{Q.\tilde{N}}$ and $K\oplus{K.\tilde{N}}$.
\end{definition}

According to the above definition, the selector $\mathcal{S}(\cdot)$ is expected to realize the following two functionalities. (1) \textbf{Ranking}: which identifies the relative importance of the neighbors; (2) \textbf{Truncation}: which determines how many top-ranked neighbors should be selected and utilized. Insufficient selection will be less informative, while excessive selection will introduce noise.

Besides, the selector is desirable for satisfying two properties. \textbf{Ad-hoc}: instead of assigning a fixed set of neighbors to each document, the selection needs to be dynamically dependent on the given counterpart: the query's neighbors are selected w.r.t. the key to be matched, and vice versa. \textbf{Lightweight}: the neighbor selector needs to be highly efficient, such that it could traverse all neighbors for the optimal subset with affordable computation overhead.

\begin{figure*}
    \centering
    \includegraphics[width=0.85\linewidth]{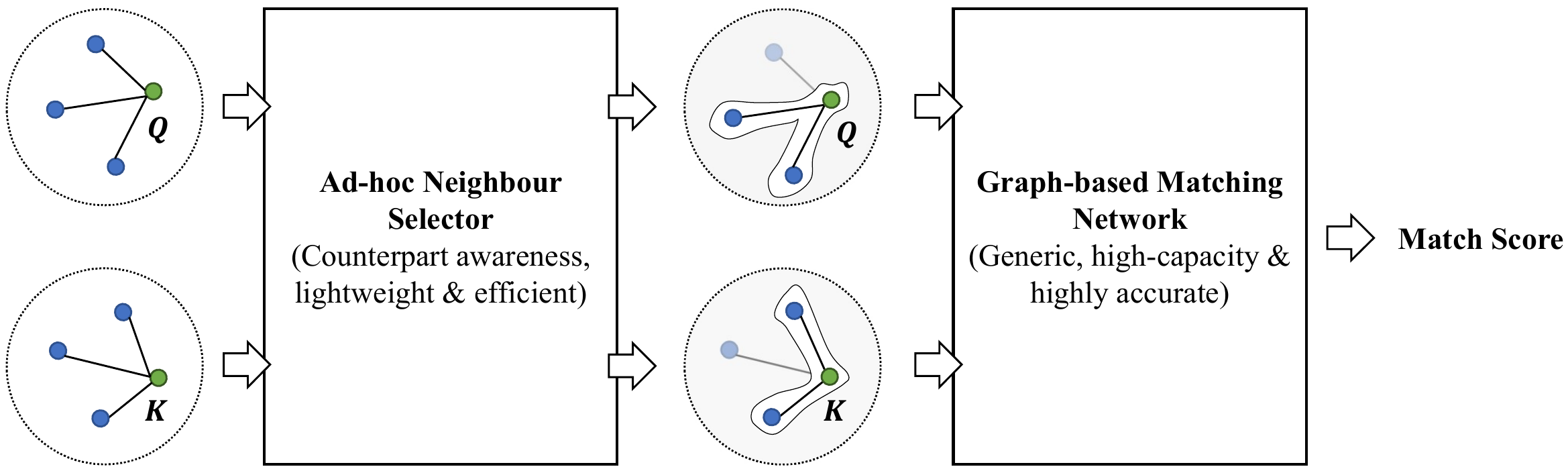}
    \caption{The whole architecture of our CDSM framework with a two-stage workflow: (1) A lightweight ad-hoc neighbor selector is deployed to filter the optimal subset of neighbors for the matching documents; (2) A high-capacity matching model is employed to calculate the relevance score with these selected neighbors.}
    \label{fig:architecture}
\end{figure*}

\section{Cascaded Deep Semantic Matching}\label{sec:framework}
In this work, we propose Cascaded Deep Semantic Matching (CDSM), a generic framework for accurate and efficient semantic matching on textual graphs. It consists of two modules: the lightweight ad-hoc neighbor selector, and the high-capacity graph-based semantic matching network. Besides, we design a hybrid optimization algorithm to train the above two modules. We also demonstrate a prototype implementation of CDSM, which realizes the properties required in Section~\ref{sec:problem_definition} and shows strong empirical performance.

\subsection{The CDSM Framework}\label{subsec:cdsm_framework}
The proposed CDSM framework is illustrated in Figure~\ref{fig:architecture}, which completes the graph-based semantic matching in two consecutive steps. First, a lightweight ad-hoc neighbor selector is developed to filter the optimal subset of useful neighbors that would help to discriminate the semantic relationship between the query and key. Second, a high-capacity graph-based matching network is employed to calculate the relevance score based on the given documents and their selected neighbors. The detailed formulations are introduced as follows.

\subsubsection{Ad-hoc Neighbor Selector} In the first stage, the CDSM takes the query document $Q$, key document $K$ and all their neighbors $Q.N=[Q.N_1,\ldots,Q.N_n]$, $K.N=[K.N_1,\ldots,K.N_n]$ as the input. Then, the ad-hoc neighbor selection is performed with two operations: 1) ranking, which estimates and compares the importance of all the neighbors, and 2) truncation, which determines how many top-ranked neighbors should be selected (i.e, the value of $k$ when making Top-$k$ selection).  

$\bullet$ \textbf{Ranking function}. We discuss two optional forms of the ranking function. The first one is the \textit{one-step} ranking function $\mathcal{R}^{one}(\cdot)$, where the importance of a query's neighbor $Q.N_i$ and a key's neighbor $K.N_j$ is computed as follows:
\begin{equation}
\begin{aligned}
    {r}(Q.N_i) &= \mathcal{R}^{one}(Q.N_i|Q,K); \\
    {r}(K.N_j) &= \mathcal{R}^{one}(K.N_j|Q,K).
\end{aligned}
\end{equation}
Because the computation of the neighbor's usefulness ${r}(Q.N_i)$ and ${r}(K.N_j)$ purely relies on the $Q$ and $K$, the importance scores of all neighbors $Q.N$ and $K.N$ can be simultaneously computed in one step. The top-$k$ neighbors can be directly selected thereafter:
\begin{equation}
\begin{aligned}
    Q.\tilde{N} = \text{top-}k({r}(Q.N_i)|\forall Q.N_i \in Q.N); \\
    K.\tilde{N} = \text{top-}k({r}(K.N_j)|\forall K.N_j \in K.N).
\end{aligned}
\end{equation}
The other option is the \textit{multi-step} ranking function, which consecutively filters useful neighbors one by one in multiple steps. In this method, the importance score of a neighbor is estimated based on not only the matching documents $Q$ and $K$, but also the already selected neighbors $Q.\tilde{N}$ and $K.\tilde{N}$ (initialized to be empty at the beginning):
\begin{equation}
\begin{aligned}
    {r}(Q.N_i) = \mathcal{R}^{mul}(Q.N_i|Q,K,Q.\tilde{N}); \\
    {r}(K.N_j) = \mathcal{R}^{mul}(K.N_j|Q,K,K.\tilde{N}).
\end{aligned}
\end{equation}
The underlying intuition is that the selected neighbors are desired of providing comprehensive information about $Q$ and $K$'s relationship. As a result, their underlying semantics are desired to be diversified. In other words, it is unnecessary to choose multiple identical or highly similar neighbors, because no additional information can be introduced by them. In this place, the selection is performed for $k$ consecutive rounds w.r.t to the already selected neighbors. The neighbor with the largest information gain is selected in each step:
\begin{equation}
\begin{aligned}
Q.\tilde{N} \leftarrow Q.\tilde{N}+ Q.N_*, ~ & Q.N \leftarrow Q.N \setminus Q.N_*, ~ & \text{where}~
Q.N_* = \mathrm{argmax}(\mathrm{r}(Q.N_i));\\
K.\tilde{N} \leftarrow K.\tilde{N}+ K.N_*, ~ & K.N \leftarrow K.N \setminus K.N_*, ~ & \text{where}~
K.N_* = \mathrm{argmax}(\mathrm{r}(K.N_i)).
\end{aligned}
\end{equation}

$\bullet$ \textbf{Truncation}. The ranking function only measures the relative importance of all neighbors, whereas it is still unclear how many top-ranked neighbors should be selected. Some documents may have very few useful neighbors for the given semantic matching task, while others may have a large number of useful neighbors. In this paper, we propose the following heuristic strategies to determine how to truncate the top-ranked neighbors to get the optimal subset, including basically static methods and more flexibly dynamic methods. All these strategies can be easily applied together with the ranking functions. In practice, the truncation strategy can be chosen based on empirical performance. 

\textit{Fixed Capacity}. The simplest way of truncation is to set a fixed capacity and always select the same number of neighbors for all documents when making the Top-$k$ selection. As discussed above, a fixed capacity may lack the flexibility to deal with documents with different numbers of useful neighbors. Thus, more complicated strategies are further introduced for a complement.

\textit{Absolute Value as Threshold}. Since the utility of neighbors is reflected by the predicted score $r(Q.N_i)$ or $r(K.N_j)$, we think there should be an absolute score threshold $\tau$ to distinguish whether a neighbor is necessary to be selected. In this case, a neighbor $Q.N_i$ will be selected if $r(Q.N_i)>\tau$, the same for $K.N_j$. Both approaches set a static threshold (capacity or value) for truncation, more flexibly dynamic methods are listed in the next.

\textit{Overall Ranking}. Recall that we define the semantic matching task in a typical ranking style as Definition~\ref{def:1}, we think the truly positive key documents would have more useful neighbors to support their connection with the query document than those negative ones. Thus, when measuring the utility of all keys' neighbors with the ranking function, the positive key counterpart would have more neighbors that achieve high ranking scores, and more neighbors should be selected for it. Based on this hypothesis, we propose to rank the neighbors of all candidate keys in a unified list, and select those neighbors within the Top-$p$ position. We illustrate the matching tasks between a query document $Q$ and a series of key documents $K_1,K_2,\ldots$ as an example. We copy $Q$ for many times to construct document pairs with all the keys, obtaining $[Q_1,K_1],[Q_2,K_2],\ldots$ where all $Q_i$ are the same. For all keys, we rank all their neighbors $K_i.N_j$ in a unified list based on the ranking score $r(K_i.N_j)$ and selected those neighbors within the Top-$p$ position for the corresponding key. With regard to the copied queries $Q_1,Q_2,\ldots$, we rank all their neighbors $Q_i.N_j$ in an unified list on top of their ranking scores $r(Q_i.N_j)$ and select the Top-$p$ neighbors. Although the $Q$ and $Q.N$ are copied, when dealing with different key documents, the ranking score of the query neighbors is decided by different keys to highlight different neighbors.

\textit{Relevance Score as Threshold}. Another intuition is that the introduction of neighbor features should strengthen the correlation between $Q$ and $K$, as presented in Definition~\ref{def:1}. Therefore, we expect that the selected neighbors can outscore the original relevance between $Q$ and $K$. At first, we measure the $Q$ and $K$'s similarity as $sim(Q,K)$ (sample implementation of $sim(Q,K)$ will be introduced in Subsection \ref{subsec:implement}). Then, a neighbor $Q.N_i/K.N_i$ will be selected if it satisfies $r(Q.N_i)>sim(Q,K)/r(K.N_i)>sim(Q,K)$.

To summarize, the method of fixed capacity selects the same number of neighbors for all documents, while the other three methods are adaptive to different matching tasks. The two kinds of strategies can be combined together to benefit each other.

\subsubsection{Graph-based Matching Network}
In the second stage, a graph-based semantic matching network is deployed. The matching network will take the query $Q$, key $K$, and the selected neighbors $Q.\tilde{N}$ and $K.\tilde{N}$ as the input. Then, it predicts $Q$ and $K$'s relevance score as:
\begin{equation}\label{eq:match}
    m(Q,K) = \mathcal{M}(Q\oplus{Q.\tilde{N}},K\oplus{K.\tilde{N}}),
\end{equation}
where ``$\oplus$'' is the graph aggregator. Though CDSM is generic and adaptive to various matching models, we desire a high-capacity graph-based semantic matching network in order to derive fine-grained relevance prediction, e.g., TextGNN and GraphFormers. (This is different from the neighbor selector which requires lightweight computation.) Besides, knowing that the matching network only needs to deal with a small set of well-selected neighbors, $|Q.\tilde{N}| \ll |Q.N|$ and $|K.\tilde{N}| \ll |K.N|$, the relevance prediction is ensured to be made with much less computation overhead.

\begin{figure*}
    \centering
    \includegraphics[width=0.9\linewidth]{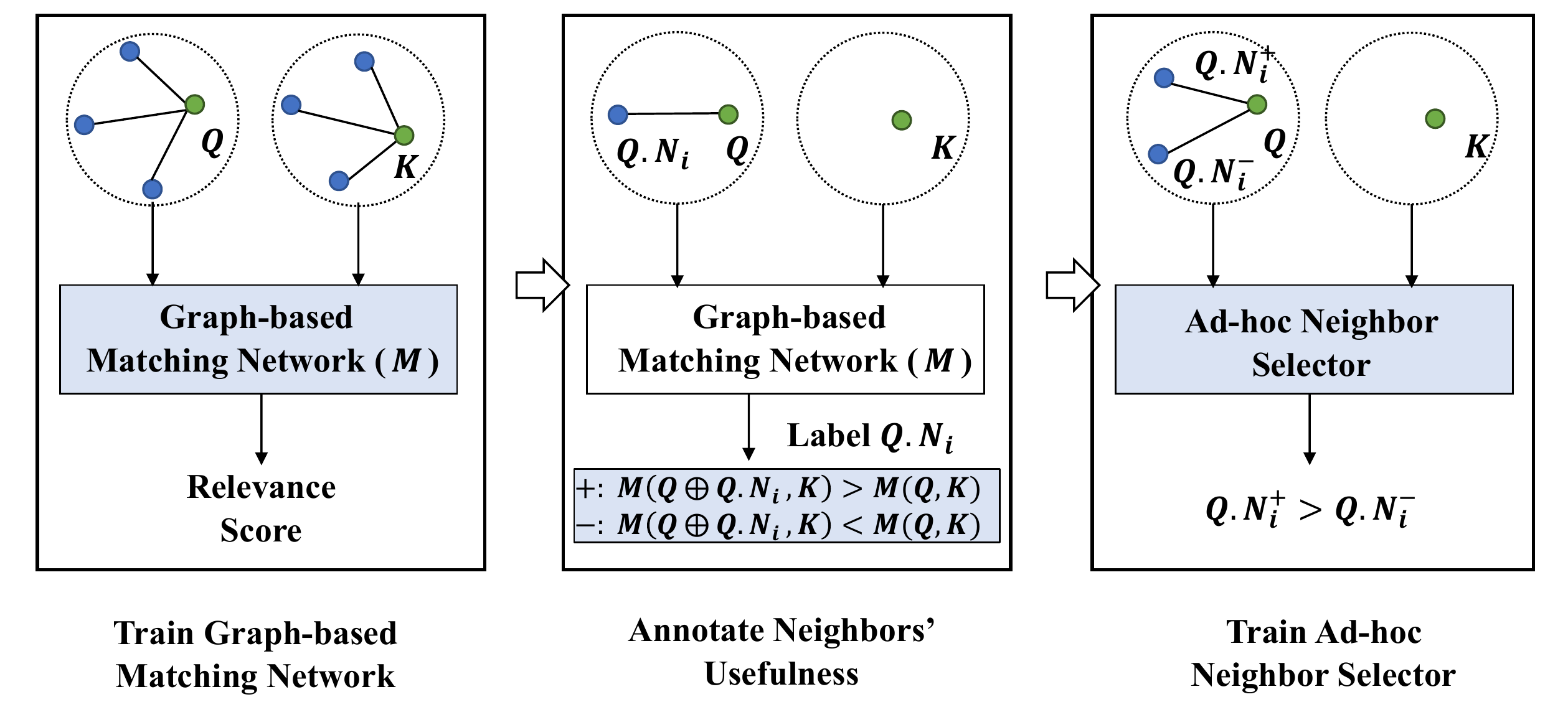}
    \caption{The hybrid optimization workflow of the CDSM framework: (1) Training the graph-based matching network $\mathcal{M}$ based on the document pairs with heuristically sampled neighbors; (2) Annotating the neighbors' usefulness with the well-trained matching network; (3) Training the neighbor selector with these weak annotations.}
    \label{fig:optimization}
\end{figure*}

\subsection{The CDSM Optimization}
The optimization of CDSM is challenging because there are no explicit measurements of neighbors' usefulness. To deal with this challenge, we propose a hybrid optimization workflow, where the neighbor selector is trained on top of weak annotations from the semantic matching network. 

$\bullet$ \textbf{Hybrid Optimization} The proposed hybrid optimization works in the following three steps, as illustrated in Figure~\ref{fig:optimization}. First, we train the graph-based matching network $\mathcal{M}$ based on the labeled document pairs and randomly sampled neighbors. Second, the well-trained semantic matching model is used to annotate the neighbor's usefulness based on Definition~\ref{def:1}. Finally, the ad-hoc neighbor selector is trained on the annotation results through contrastive learning. The details of each step are illustrated in the following. 

$\bullet$ \textbf{Train Semantic Matching Network}. Given a positive pair of documents $Q, K$ and their neighbors $Q.N, K.N$ which are selected by random sampling or heuristic rules, the matching score is calculated as $m(Q,K) = \mathcal{M}(Q\oplus Q.N, K\oplus K.N)$ (Eq.~\ref{eq:match}). Then, we randomly sample a batch of negative key documents $[K_1^-,K_2^-,\ldots,K_M^-]$. The matching network is trained to distinguish the positive keys from the negative keys by minimizing the classification loss.
\begin{equation}
    \mathcal{L}^{cls} = -\frac{\exp(m(Q,K^+))}{\exp(m(Q,K^+))+\sum_{i}\exp(m(Q,K_i^-))}.
\end{equation}
In our implementation, we make use of ``in-batch negative samples''~\cite{luan2020sparse,Karpukhin2020inbatch} to reduce the computation cost, where for one positive document pair, the keys of the other samples in the same mini-batch are viewed as negative keys.

$\bullet$ \textbf{Train Ad-hoc Neighbor Selector}. The neighbor selector is trained via weak supervision, with neighbors’ usefulness annotated by the semantic matching model trained in the last step. Based on whether the one-step or multi-step ranking function is utilized, the data annotation is performed in two different ways.  

When the one-step ranking function $R^{one}(\cdot)$ is utilized, a neighbor's usefulness is determined purely based on the input features of $Q$ and $K$. For a positive pair of documents $Q, K$, their semantic matching score without any neighbor features can be calculated as $\mathcal{M}(Q,K)$. Then, the usefulness of each query neighbor $Q.N_i$ for this matching task is evaluated as $\mathcal{M}(Q\oplus Q.N_i,K)$, and that of each key neighbor $K.N_i$ is denoted as $\mathcal{M}(Q,K\oplus K.N_i)$. A neighbor is useful if it improves the matching network's prediction about the relationship between $Q$ and $K$, based on which the labels are generated by the formulations:
\begin{equation}
\begin{split}
l(Q.N_i) = 
\begin{cases}
+: ~ \mathcal{M}(Q\oplus{Q.N_i}, K) > \mathcal{M}(Q, K),
\\ 
-: ~ \mathcal{M}(Q\oplus{Q.N_i}, K) \leq \mathcal{M}(Q,K),
\end{cases}
\\
l(K.N_i) = 
\begin{cases}
+: ~ \mathcal{M}(Q, K\oplus{K.N_i}) > \mathcal{M}(Q, K),
\\ 
-: ~ \mathcal{M}(Q, K\oplus{K.N_i}) \leq \mathcal{M}(Q, K).
\end{cases}
\end{split}
\end{equation}
$l(Q.N_i)$ and $l(K.N_i)$ are the labels of the neighbors, and $\mathcal{M}(\cdot)$ denotes the matching network's predicting function.

When the multi-step ranking function $R^{mul}(\cdot)$ is utilized, the neighbor's usefulness is measured based on not only $Q$ and $K$, but also the already selected neighbors. As a result, we define a neighbor to be useful if it further improves the matching network's prediction about $Q$ and $K$'s relevance after introducing it into the currently selected neighbor subset $Q.\tilde{N}, K.\tilde{N}$.
\begin{equation}
\begin{split}
l(Q.N_i) = 
\begin{cases}
+: ~ \mathcal{M}(Q\oplus\{Q.N_i,Q.\tilde{N}\}, K) > \mathcal{M}(Q\oplus{Q.\tilde{N}}, K),
\\ 
-: ~ \mathcal{M}(Q\oplus\{Q.N_i,Q.\tilde{N}\}, K) \leq \mathcal{M}(Q\oplus{Q.\tilde{N}}, K),
\end{cases}
\\
l(K.N_i) = 
\begin{cases}
+: ~ \mathcal{M}(Q, K\oplus\{K.N_i,K.\tilde{N}\}) > \mathcal{M}(Q, K\oplus{K.\tilde{N}}),
\\ 
-: ~ \mathcal{M}(Q, K\oplus\{K.N_i,K.\tilde{N}\}) \leq \mathcal{M}(Q, K\oplus{K.\tilde{N}}).
\end{cases}
\end{split} 
\end{equation}

With the neighbors of all positive document pairs annotated, we construct neighbor pairs comprised of a positive neighbor $N_i^+$ and a negative neighbor $N_i^-$ to train the selector by contrastive learning. Taking a given query document $Q$ as an example, the neighbor selector is learned by minimizing the following loss:
\begin{equation}
\mathcal{L}^{sel} = -\frac{1}{1+\exp(-(r(Q.N_i^+)-r(Q.N_i^-)))}.
\end{equation}
$r(Q.N_i^+)$ and $r(Q.N_j^-)$ are calculated by the corresponding $R^{one}(\cdot)$ or $R^{mul}(\cdot)$, when different ranking functions are applied.

\subsection{The CDSM Implementation}\label{subsec:implement}
In this place, we show a prototype implementation of the ad-hoc neighbor selector, which will be used in our experiments. In order to guarantee efficiency, we adopt a lightweight architecture, including a CNN-based text encoder and a ranking function. The implementation of graph-based semantic matching networks follows baseline models, whose details will be introduced in Section~\ref{sec:experimental_setting}.

As for the CNN-based text encoder, the input is an individual text, taking the query $Q=[w_1^Q,w_2^Q,\ldots]$ as an example. The first layer is a word embedding layer that converts tokens into low-dimensional vectors. By passing the query through this layer, we obtain a word embedding matrix $V^Q=[v_1^Q,v_2^Q,\ldots]$. The second layer is a 1-d CNN layer, capturing local context information within the text sequence to obtain better word representations. As for the i-th term, its context-aware representation $c_i^Q$ is calculated as:
\begin{align}
    c_i^Q = \text{ReLU}(F_w\times v_{(i-k):(i+k)}^Q + b_w),
\end{align}
where $v_{(i-k):(i+k)}^Q$ means the word embeddings from the position $(i-k)$ to $(i+k)$. $2k+1$ is the size of the context window. $F_w$ and $b_w$ are the parameters of CNN filters. Through the 1-d CNN layer, the output is a matrix of contextual word representations, denoted as $C^Q=[c_1^Q,c_2^Q,\ldots]$. Considering that different words in a sentence contribute different informativeness, we set the third layer as a word-level attention network. The query in the attention mechanism is a trainable dense vector $q_w$. We compute the attention weight of each word based on the interaction between the query $q_w$ and the context-aware word representation, i.e.,
\begin{align}
    \alpha _i = \frac{\exp(\alpha_i)}{\sum_{j=1}^{|Q|}\exp(\alpha_j)},\quad \alpha _i = (c_i^Q)^T \text{tanh}(W_q\times q_w + b_q).
\end{align}
Finally, the text representation $r^Q$ of $Q$ is the weighted sum of all word representations based on the attention weights, as:
\begin{align}
    r^Q = \sum_{i=1}^{|Q|}\alpha_ic^Q_i.
\end{align}
Using this CNN-based text encoder, we are able to obtain the text representation for the matching documents $Q, K$ and all the presented neighbors.

With $Q, K$ and all neighbors represented as vectors, the ranking function is employed on them to evaluate the neighbors' importance. With regard to the one-step ranking function, we calculate the usefulness of a neighbor as the relevance between it and the matching counterpart:
\begin{align}
    \mathcal{R}^{one}(Q.N_i|Q,K) = (r^{Q.N_i})^T r^{K},\\
    \mathcal{R}^{one}(K.N_i|Q,K) = (r^{K.N_i})^T r^{Q}.
\end{align}

As for the multi-step ranking function, it calculates the usefulness for a neighbor $Q.N_i/K.N_i$ according to both the selected neighbor set $Q.\tilde{N}/K.\tilde{N}$ and the matching documents $Q, K$. To highlight the different aspects of information that the already selected neighbors cover, we perform max-pooling along the last dimension to aggregate all selected neighbors. Thus, the usefulness is calculated as:
\begin{align}
    \mathcal{R}^{mul}(Q.N_i|Q,K,Q.\tilde{N}) = \varphi([r^Q,\text{max-pool}(r^{N_i}|\forall N_i \in N)])^T r^K,
\end{align}
where $N = Q.\tilde{N} + Q.N_i$. $\varphi(\cdot)$ is an MLP layer, $[\cdot,\cdot]$ indicates vector concatenation, and $\text{max-pool}(\cdot)$ is the max-pooling operation along the last dimension.

For the function $sim(\cdot)$ to measure $Q$ and $K$'s similarity, it works as:
\begin{align}
    sim(Q,K) = (r^Q)^T r^K.
\end{align}

\begin{table}[]
    \centering
    \caption{Time complexity analysis of CDSM and other graph-based matching methods.}
    \begin{tabular}{c|ccc}
    \toprule
        Method & All Neighbors & Heuristic Selection & CDSM \\
    \hline
        Time Complexity & $n * T_m$ & $n * T_h + k * T_m$ & $n * T_s + k * T_m$ \\
    \bottomrule
    \end{tabular}
    \label{tab:time_complexity}
\end{table}
\subsection{The CDSM Efficiency Analysis}
As stated in Section~\ref{sec:intro}, filtering a set of irrelevant neighbors with a lightweight neighbor selector could improve both the efficiency and accuracy of graph-based semantic matching. Here, we deploy a mathematical time complexity analysis to prove the CDSM efficiency. We compare the computation time for the semantic matching task of directly using all neighbors, sampling neighbors heuristically and selecting neighbors with our CDSM framework. The comparison results are displayed in Table~\ref{tab:time_complexity}.

In Table~\ref{tab:time_complexity}, $T_m$ represents the time cost of a high-capacity semantic matching network to encode a document. $T_s$ indicates the time cost of the lightweight selector to process a neighbor node ($T_s << T_m$). $T_h$ means the time cost of selecting a neighbor through heuristic methods, such as random sampling. $n$ and $k$ represent the number of all available neighbors and that of the selected neighbors respectively. As such, when $k << n$, CDSM saves a lot of time for computing the fine-grained document representations compared with using all neighbors. Our designed lightweight selector obtains document representations through CNN, thus the time cost $T_s$ would be so small that CDSM can be comparably efficient as the heuristic sampling methods.

In the next sections, we conduct experiments to verify the strong performance and efficiency of the above introduced simplified selector.

\begin{table}[]
    \centering
    \caption{Statistics of the three datasets.}
    \begin{tabular}{p{1.8cm}|p{2.0cm}|p{1.6cm}|p{1.6cm}}
        \toprule
         Dataset & \textbf{Wikidata5M} & \textbf{DBLP} & \textbf{Product}  \\
        \hline
        \#Item & 4,018,299 & 3,691,796 & 2,049,487\\
        Avg.\#N & 27.35 & 46.94 & 17.02\\
        \#Train & 7,145,834 & 3,009,506 & 3,004,199 \\
        \#Valid & 66,167 & 60,000 & 50,000\\
        \#Test & 100,000 & 100,000 & 536,575\\
        \bottomrule
    \end{tabular}
    \label{tab:data}
\end{table}

\section{Experimental Settings}\label{sec:experimental_setting}
\subsection{Datasets and Evaluations}
\subsubsection{Datasets}
We use three large-scale textual graph datasets for evaluation. 1) \textbf{Wikidata5M}\footnote{https://deepgraphlearning.github.io/project/wikidata5m}, a million-scale knowledge graph dataset introduced in~\cite{Wang2021Wikidata}. It contains entities and their connections. Each entity has an aligned passage from the corresponding Wikipedia page and we take the first sentence from the passage as its textual description. 2) \textbf{DBLP}\footnote{https://originalstatic.aminer.cn/misc/dblp.v12.7z}. This dataset contains academic papers up to 2020-04-09 collected from DBLP\footnote{https://dblp.uni-trier.de/}. The connections between different papers are built upon their citation relationships. We use each paper's title as its textual description. 3) \textbf{Product}. This is a product graph dataset constructed from the real-world Bing search engine. We track each user's web browsing events along with the targeted product web pages and divide her continuous events into sessions with 30 minutes of user inactivity as interval~\cite{yao2020peps}. Following a typical product graph construction method in e-commerce platforms~\cite{wang2018billion}, the products within a common session are linked to each other. Each product has its corresponding textual description which indicates the product name, brand, type and so on.

For these three datasets, we uniformly sample at most 50 neighbors for each center node from the set comprised of its one-order and two-order neighbors. The statistics of datasets are listed in Table~\ref{tab:data}.

\subsubsection{Evaluations}
In this paper, we define the graph-based semantic matching task in the form of a ranking problem, as Definition~\ref{def:2}. For each positive pair of query and key document, we sample 29 negative key documents, and the target is to rank the positive key higher than these negative keys. Two common ranking metrics are leveraged to evaluate the matching accuracy: \textbf{precision@1 (p@1)} and \textbf{ndcg}.

\subsection{Baselines}
Our proposed CDSM is a generic framework to improve both the accuracy and efficiency of the graph-based semantic matching task. It is highlighted for the two-stage workflow instead of a specific semantic matching model: a lightweight neighbor selector to obtain informative neighbors and then the graph-based semantic matching task is conducted, where various mainstream graph-based matching models can be adapted. 

\subsubsection{Graph-based Semantic Matching Models}
To verify its generality, various mainstream graph-based semantic matching models are incorporated as the backbone of the matching network. In these models, texts are all encoded with BERT. The last layer's [CLS] token embedding is treated as the document representation.

$\bullet$ \textbf{BERT}~\cite{Devlin2019BERT}: It calculates the semantic matching score as the dot product between the query's and key's [CLS] vectors encoded by BERT respectively, without any neighbor features.

$\bullet$ \textbf{TextGNN}~\cite{Zhu2021TextGNN}: This is one of the most recent graph-based semantic matching methods, with pre-trained language model BERT as the text encoder and GNNs to aggregate the information from neighbors. We consider the following forms of GNNs. \textbf{GAT}~\cite{Velickovic2017GAT}, which combines neighbors and the center document as a weighted sum of all their text vectors. The weight of each text vector is calculated as the attention score with the center document. \textbf{GraphSage}~\cite{Hamilton2017GraphSage}, where the neighbors are first aggregated, then the aggregation result is concatenated with the center document embedding and passed through a dense layer to generate the final representation for semantic matching. There are four different aggregation variants: \textbf{MaxSage} and \textbf{MeanSage}, which aggregate neighbors by max-pooling and mean-pooling respectively. \textbf{AttnSage} aggregates the neighbors based on their attention scores with the center document. \textbf{Counterpart} aggregates the neighbors based on their attention scores with the matching counterpart, as presented in Section~\ref{sec:problem_definition}. The above variants of TextGNN are denoted as \textbf{TextGNN(GAT/Max/Mean/Attn/CP)} in our experiments.

$\bullet$ \textbf{BERT4Graph}: This is a naive extension of BERT from traditional text matching to graph-based semantic matching. For the query, it concatenates the tokens of the query and all neighbors, with a [CLS] token added to the head. Then, the joint token sequence is inputted into BERT for fine-grained interactions. The last layer's [CLS] token embedding is used for the semantic matching, the same for the key document. The semantic matching score is calculated as their cosine similarity.

$\bullet$ \textbf{GraphFormers}~\cite{Yang2021GraphFormer}: This model adopts a deeper fusion of GNNs and text encoders than TextGNN to make more effective use of neighbor features. It nests GNN components between the transformer layers to conduct iterative text encoding and neighbor aggregation.

Except that BERT uses no neighbor information, the other matching models consider neighbors. We apply them as the backbone of the matching network, and compare their performance under different neighbor selection methods.

\subsubsection{Selection Approaches}
Furthermore, to verify the effectiveness of the ad-hoc neighbor selector under our CDSM framework, we compare it with other selection approaches, including random sampling and two heuristic rule-based methods. All these baselines assign a static set of neighbors to each document, without dynamically considering the matching counterparts like our CDSM. Details are listed as follows.

$\bullet$ \textbf{Random Sampling}: For each document, a set of distinct neighbors is randomly sampled from all the presented neighbors.

$\bullet$ \textbf{Popularity}: For each document, we rank all its neighbors according to their popularity, i.e., how many nodes they have connections with. Then, the top-$k$ popular neighbors are selected.

$\bullet$ \textbf{Similarity}: For each center document, the top-$k$ neighbors that are most similar to it are selected. The similarity is computed as the dot product between their vectors encoded by BERT. 

$\bullet$ \textbf{CDSM}: This indicates the ad-hoc neighbor selector under our framework.

\begin{table*}[t]
    \centering
    \setlength{\abovecaptionskip}{0.1cm}
    \setlength{\belowcaptionskip}{0.1cm}
    \caption{Overall performance of different graph-based semantic matching models with different available neighbor sets: all neighbors, randomly sampled 5 neighbors, 5 neighbors selected according to the popularity, the similarity with the center document and the usefulness measured by CDSM. ``$\dagger$'' denotes that the result is significantly better than other selection methods (except for all neighbors) in t-test with $p \textless 0.05$ level. ``$^*$'' denotes that the result is significantly better than that with all neighbors in t-test with $p \textless 0.05$ level. Results of the best neighbor selection method are shown in bold. }
    \label{tab:overall_performance}
    \begin{tabular}{p{0.07\textwidth}p{0.17\textwidth}p{0.04\textwidth}p{0.04\textwidth}p{0.04\textwidth}p{0.04\textwidth}p{0.04\textwidth}p{0.04\textwidth}p{0.04\textwidth}p{0.04\textwidth}p{0.045\textwidth}p{0.04\textwidth}}
        \toprule
        \multirow{2}*{Data} & \multirow{2}*{Methods} & \multicolumn{2}{c}{All} & \multicolumn{2}{c}{Random (5)} & \multicolumn{2}{c}{Popularity (5)} & \multicolumn{2}{c}{Similarity (5)} & \multicolumn{2}{c}{CDSM (5)} \\ 
        \cmidrule(lr){3-4}
        \cmidrule(lr){5-6}
        \cmidrule(lr){7-8}
        \cmidrule(lr){9-10}
        \cmidrule(lr){11-12}
         & & p@1 & ndcg & p@1 & ndcg & p@1 & ndcg & p@1 & ndcg & p@1 & ndcg\\
         \hline
         \multirow{9}*{Wiki} & BERT & .644 & .817 &  - &  - &  - &  - &  - & - &  - &  - \\
         & TextGNN(GAT) & .580 & .772 & .521 & .732 & .416 & .664 & \textbf{.587} & \textbf{.781} & .581 & .778 \\
         & TextGNN(Mean) & .694 & .850 & .678 & .840 & .627 & .810 & .662 & .828 & \textbf{.694}$\dagger$ & \textbf{.850}$\dagger$ \\
         & TextGNN(Max) & .687 & .844 & .676 & .839 & .653 & .827 & .666 & .831 & \textbf{.697}$\dagger$ & \textbf{.852}$\dagger$ \\
         & TextGNN(Attn) & .696 & .850 & .679 & .840 & .620 & .803 & .664 & .829 & \textbf{.694}$\dagger$ & \textbf{.850} \\
         & TextGNN(CP) & .705 & .856 & .681 & .841 & .679 & .843 & .674 & .836 & \textbf{.704}$\dagger$ & \textbf{.855}$\dagger$ \\
         & BERT4Graph & - &  - & .690 & .846 & .622 & .789 & .683 & .841 & \textbf{.705}$\dagger$ & \textbf{.858}$\dagger$ \\
         & GraphFormers & .704 & .854 & .689 & .845 & .698 & .852 & .685 & .841 & \textbf{.707} & \textbf{.860} \\
         \hline
         \hline
         \multirow{9}*{DBLP} & BERT & .821 & .914 &  - &  - &  - &  - &  - &  - &  - &  -  \\
         & TextGNN(GAT) & .826 & .918 & .776 & .892 & .790 & .900 & .851 & .929 & \textbf{.860}$\dagger^*$ & \textbf{.936}$^*$ \\
         & TextGNN(Mean) & .845 & .908 & .817 & .894 & .816 & .894 & .815 & .891 & \textbf{.844}$\dagger$ & \textbf{.907}$\dagger$ \\
         & TextGNN(Max) & .805 & .888 & .812 & .891 & .813 & .892 & .812 & .890 & \textbf{.833}$\dagger$$^*$ & \textbf{.902}$\dagger$$^*$ \\
         & TextGNN(Attn) & .843 & .906 & .815 & .892 & .813 & .892 & .813 & .875 & \textbf{.842}$\dagger$  & \textbf{.906}$\dagger$ \\
         & TextGNN(CP) & .852 & .917 & .819 & .901 & .818 & .901 & .816 & .899 & \textbf{.850}$\dagger$ & \textbf{.916}$\dagger$ \\
         & BERT4Graph &  - &  - & .864 & .936 & .873 & .941 & .878 & .942 & \textbf{.901}$\dagger$ & \textbf{.955}$\dagger$ \\
         & GraphFormers & .903 & .955 & .877 & .944 & .882 & .946 & .880 & .942 & \textbf{.909}$\dagger$ & \textbf{.959}$\dagger$ \\
         \hline
         \hline
         \multirow{9}*{Product} & BERT & .481 & .707 &  - &  - &  - &  - &  - &  - &  - &  - \\
         & TextGNN(GAT) & .568 & .774 & .578 & .781 & .545 & .760 & .567 &  .770 & \textbf{.588}$^*$ & \textbf{.787} \\
         & TextGNN(Mean) & .746 & .883 & .723 & .868 & .721 & .867 & .703 & .852 & \textbf{.748}$\dagger$ & \textbf{.885}$\dagger$ \\
         & TextGNN(Max) & .730 & .873 & .728 & .871 & .721 & .867 & .712 & .858 & \textbf{.756}$\dagger$$^*$ & \textbf{.890}$\dagger$$^*$ \\
         & TextGNN(Attn) & .744 & .882 & .725 & .869 & .722 & .867 & .706 & .847 & \textbf{.741}$\dagger$ & \textbf{.881} \\
         & TextGNN(CP) & .756 & .889 & .728 & .871 & .723 & .867 & .710 & .857 & \textbf{.760}$\dagger$ & \textbf{.892}$\dagger$ \\
         & BERT4Graph & - &  - & .739 & .877 & .733 & .873 & .722 & .864 & \textbf{.771}$\dagger$ & \textbf{.898}$\dagger$ \\
         & GraphFormers & .737 & .879 & .715 & .864 & .711 & .862 & .701 & .852 & \textbf{.759}$\dagger$$^*$ & \textbf{.891}$\dagger$ \\
        \bottomrule
    \end{tabular}
\end{table*}

\subsection{Model Settings}
In our experiments, UniLMv2-base \citep{Bao2020UniLMv2} is applied as the backbone of BERT. As for the ad-hoc neighbor selector, the word embedding layer is initialized with the underlying word vectors in UniLMv2-base. Dimensions of the word embeddings and hidden states are set as 768. We utilize the uncased WordPiece~\cite{Wu2016google} to tokenize all text sequences. The max length of sequences is set as 64 for Wikidata5M, 32 for DBLP, 32 for Product. For the CNN text encoder, the size of the context window is 3. The batch size is 300 and the learning rate is 1e-5. Adam optimizer is applied for training the selector.

As for high-capacity matching models, on Wikidata5M, DBLP and Product, the batch size is 160, 240, 240; the learning rates are 5e-6, 1e-6, 1e-5. `In-batch negative samples' and Adam are used for optimization. Each training sample is comprised of two groups of documents: 1 query document with 5 randomly sampled neighbors; and 1 key document with 5 randomly sampled neighbors. The training is conducted with 8x Nvidia V100-16GB GPUs.

\begin{figure}
    \centering
    \includegraphics[width=0.6\linewidth]{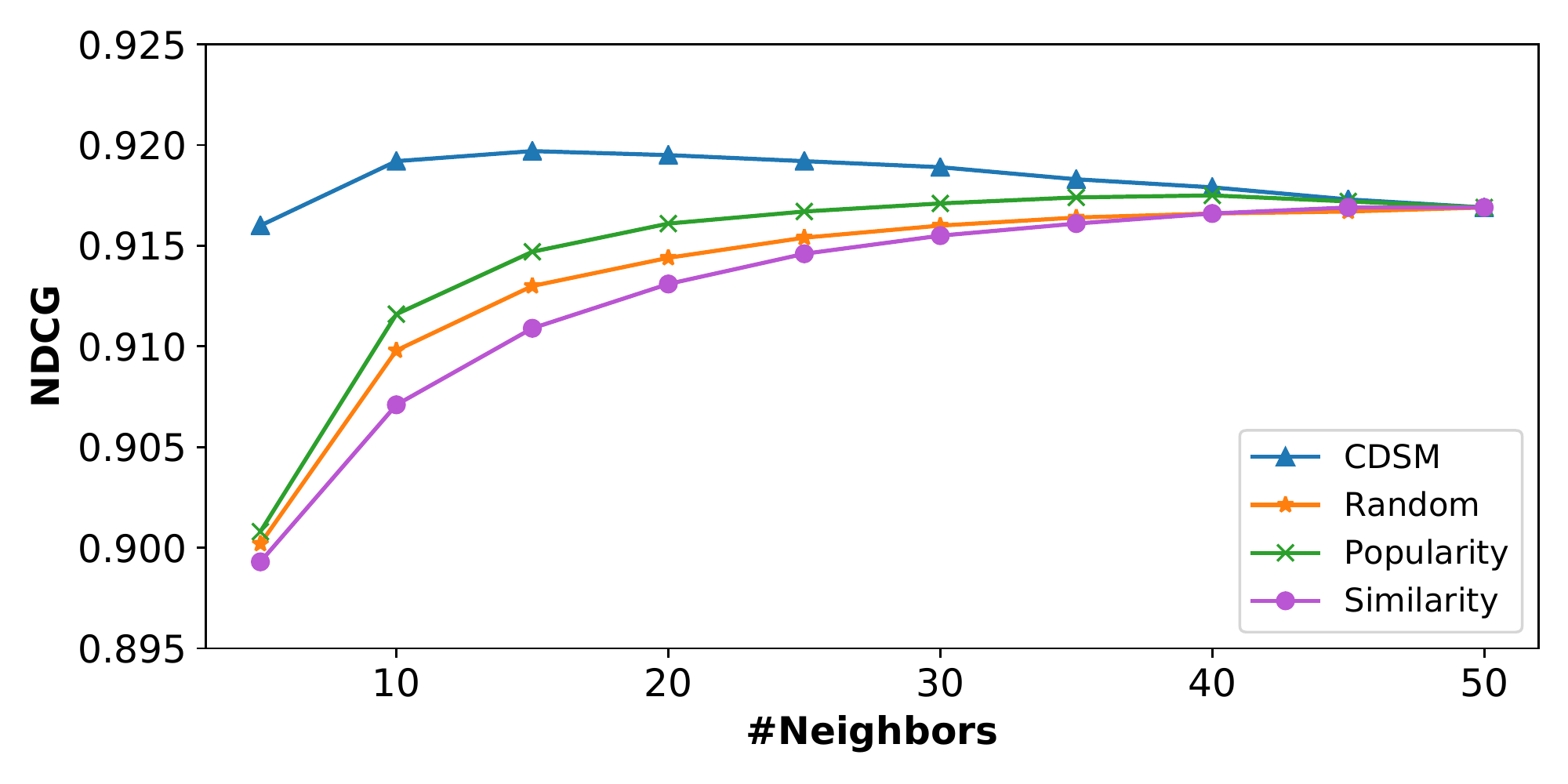}
    \caption{Performance curves of the TextGNN(CP) model with different numbers of neighbors selected by different methods.}
    \label{fig:sensitivity}
\end{figure}

\begin{figure*}
    \centering
    \includegraphics[width=1.0\linewidth]{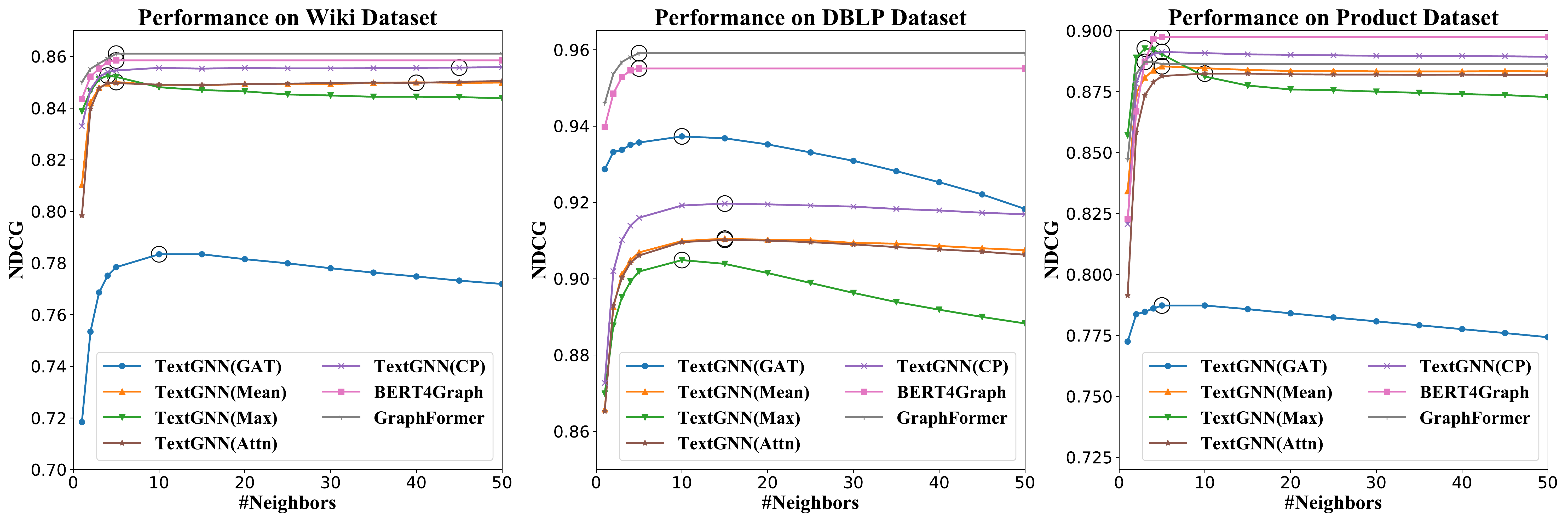}
    \caption{Performance curves of different semantic matching models with different number of neighbors selected by CDSM. The best results are marked by black circles.}
    \label{fig:performance_curve}
\end{figure*}

\section{Experimental Results and Analysis}~\label{sec:results}
In this section, we conduct extensive experiments to explore the following research questions:
\begin{itemize}[leftmargin=*]
    \item \textbf{RQ1}: As a generic framework, can CDSM consistently improve both the accuracy and efficiency of the semantic matching task when combined with various graph-based semantic matching networks? (Discussed in Subsection~\ref{subsec:rq1})
    \item \textbf{RQ2}: Different truncation approaches are proposed to determine the number of selected neighbors. How do they impact the effect and efficiency of the semantic matching task respectively? (Discussed in Subsection~\ref{subsec:rq2})
    \item \textbf{RQ3}: Can the simplified selector learn to identify the usefulness of neighbors precisely enough? (Discussed in Subsection~\ref{subsec:rq3})
    \item \textbf{RQ4}: How about the effect and efficiency of the one-step and multi-step ranking functions in CDSM? (Discussed in Subsection~\ref{subsec:rq4})
\end{itemize}

\subsection{Overall Performance (RQ1)}~\label{subsec:rq1}
To demonstrate the generality and effectiveness of CDSM, we compare the performance of different semantic matching models in various scenarios where different neighbor sets are available: all neighbors, 5 randomly sampled neighbors, 5 most popular neighbors, 5 neighbors most similar with the center document and 5 neighbors filtered by the ad-hoc neighbor selector in CDSM. The comparison results are presented in Table~\ref{tab:overall_performance}. We also conduct sensitivity testing with TextGNN(CP) on the number of neighbors selected by different methods, shown in Figure~\ref{fig:sensitivity}. Furthermore, we check the matching accuracy with the top $k$ neighbors selected by CDSM as $k$ becomes larger, shown in Figure~\ref{fig:performance_curve}. Due to the limitation of computing resources, at most 5 neighbors are tested for BERT4Graph and GraphFormers. Observing all these results, we have several findings:

First, \textbf{extrinsic neighbor features provide helpful information for the semantic matching task, but these neighbor features are indeed highly noisy. Selecting an optimal neighbor subset can achieve better performance than using all neighbors.} With joint consideration of the center document and neighbors, these graph-based semantic matching models outperform the BERT only baseline in most cases. BERT4Graph and GraphFormers which make fuller use of the neighbor features achieve the best results. However, only a small fraction of neighbors actually contribute to the semantic matching accuracy while others are noise. Comparing the results of different graph-based matching models with all neighbors and the best-performed 5 neighbors (shown in bold) on all datasets in Table~\ref{tab:overall_performance}, most matching models present similar performance. This result demonstrates that a lot of neighbors take no information to improve the matching accuracy. Focusing on the performance curves displayed in Figure~\ref{fig:performance_curve}, the curve of most models rises first and then goes down as more and more neighbors with little informativeness are involved. This trend proves that selecting a set of truly informative neighbors can achieve better matching accuracy than utilizing all neighbors.

Second, \textbf{our proposed lightweight ad-hoc neighbor selector has the ability to efficiently select the truly informative neighbors to enhance the semantic matching accuracy.} As shown in Table~\ref{tab:overall_performance}, the 5 neighbors selected by our CDSM framework perform much better than the neighbor subsets constructed by other selection baselines. Random sampling is unable to distinguish the usefulness of neighbors. In Popularity and Similarity, a static set of neighbors with a specific attribute is assigned to each document and used in matching tasks with various counterparts, thus leading to negative performance. Especially for the Popularity method, a popular neighbor node on the textual graph may be general and contain less informativeness to help specific document matching tasks, which is obvious on the Wiki dataset. Differently, our neighbor selection is performed in an ad-hoc manner to filter neighbors that are relevant to the current counterpart to facilitate the current matching task. Figure~\ref{fig:sensitivity} further presents the comparison between different selection methods with different numbers of neighbors. CDSM could always filter more informative neighbors to achieve better results than other methods, especially for cases where the computation resources are limited and the number of available neighbors is small. From Figure~\ref{fig:performance_curve}, we can come to that the neighbors ranked at high positions by CDSM significantly improve the semantic matching accuracy, whereas the neighbors with lower scores hardly improve the accuracy and even reduce the results. This observation proves that our selector can precisely measure the usefulness of neighbors for a specific matching task. Besides, we also test the time and space cost of neighbor selections. Table~\ref{tab:cost_comparison} shows that the neighbor selector is sufficiently lightweight and efficient.

Third, \textbf{our generic CDSM framework can be well combined with various graph-based matching models to consistently achieve better effects and efficiency on the semantic matching task.} As shown in Table~\ref{tab:overall_performance}, for different graph-based matching models, our CDSM framework can consistently select an effective neighbor subset to achieve comparable or better results than using all neighbors. This verifies the generality of our proposed CDSM framework. In Figure~\ref{fig:performance_curve}, we find that the best results of different models are all achieved with a small subset of informative neighbors selected by our selector, instead of encoding all the presented neighbors. Thus, a large amount of unnecessary encoding cost is saved and much background noise is avoided to get better results. Observing Table~\ref{tab:cost_comparison}, little selection cost is increased while large encoding cost can be reduced by selecting a subset of neighbors by CDSM. Therefore, both the effects and efficiency of semantic matching can be improved with the CDSM framework.

To conclude, we confirm that \textbf{neighbor selection is critical for the graph-based semantic matching task. Our CDSM framework can efficiently select a more effective neighbor subset for the matching task, improving both effectiveness and efficiency.}

\begin{table*}[t]
    \centering
    \caption{Comparison of truncation methods. ``Avg.N'' indicates the average number of selected neighbors.}
    \label{tab:stopping_conditions}
    \begin{tabular}{p{0.07\textwidth}p{0.17\textwidth}p{0.055\textwidth}p{0.05\textwidth}p{0.055\textwidth}p{0.05\textwidth}p{0.055\textwidth}p{0.05\textwidth}p{0.055\textwidth}p{0.05\textwidth}}
    \toprule
    \multirow{2}*{Datasets} & \multirow{2}*{Methods} & \multicolumn{2}{c}{Fixed Capacity} & \multicolumn{2}{c}{Absolute Score} & \multicolumn{2}{c}{Overall Ranking} & \multicolumn{2}{c}{Relevance Score}\\
    \cline{3-10}
    % \cmidrule(lr){3-4}
    % \cmidrule(lr){5-6}
    % \cmidrule(lr){7-8}
    % \cmidrule{lr}{9-10}
    & & NDCG & Avg.N & NDCG & Avg.N & NDCG & Avg.N & NDCG & Avg.N \\
    
    \hline
    \multirow{5}*{Wiki} & TextGNN(GAT) & .784 & 14.00 & .784 & 7.09 & \textbf{.785} & \textbf{4.04} & .784 & 5.79 \\
    & TextGNN(Mean) & .850 & 6.00 & .850 & 4.23 & \textbf{.851} & \textbf{5.62}  & .850 & 9.99 \\
    & TextGNN(Max) & .853 & 5.00 & .852 & 2.99 &  .852 & 2.80 & \textbf{.853} & \textbf{3.22} \\
    & TextGNN(Attn) & .850 & 7.00 & .850 & 5.71 & .850 & 5.71 & \textbf{.851} & \textbf{10.28} \\
    & TextGNN(CP) & .856 & 20.00 & .855 & 10.33 & .856 & 14.85 & \textbf{.856} & \textbf{10.66} \\
    \hline
    \hline
    \multirow{5}*{DBLP} & TextGNN(GAT) & .937 & 10.00 & \textbf{.938} & \textbf{10.13} & .938 & 11.06 & .937 & 6.80 \\
    & TextGNN(Mean) & .911 & 16.00 & .910 & 11.06 & .911 & 11.80 & \textbf{.911} & \textbf{8.13} \\
    & TextGNN(Max) & .905 & 11.00 & .901 & 8.95 & .905 & 8.47 & \textbf{.905} & \textbf{7.32} \\
    & TextGNN(Attn) & .910 & 15.00 & .910 & 12.93 & .910 & 12.47 & \textbf{.910} & \textbf{8.87} \\
    & TextGNN(CP) & .920 & 18.00 & .919 & 14.28 & .920 & 14.32 & \textbf{.920} & \textbf{9.48} \\
    \hline
    \hline
    \multirow{5}*{Product} & TextGNN(GAT) & .788 & 7.00 & \textbf{.788} & \textbf{2.27} & .788 & 3.36 & .788 & 4.77   \\
    & TextGNN(Mean) & .885 & 6.00 & .885 & 4.71 & \textbf{.886} & \textbf{3.27} & .886 & 3.91  \\
    & TextGNN(Max) & .893 & 3.00 & .893 & 2.56 &.892 & 1.76 & \textbf{.893} & \textbf{2.55} \\
    & TextGNN(Attn) & .883 & 8.00 & .883 & 5.82 & \textbf{.883} & \textbf{4.05} & .882 & 4.46 \\
    & TextGNN(CP) & .892 & 6.00 & .892 & 4.67 & \textbf{.892} & \textbf{3.31} & .891 & 4.34 \\
    \bottomrule
    \end{tabular}
\end{table*}

\subsection{Analysis of Truncation (RQ2)}~\label{subsec:rq2}
As for the ad-hoc neighbor selector in our CDSM framework, in addition to the ranking function that predicts the relative importance of all neighbors, we set truncation to determine how many top-ranked neighbors should be selected. Several different truncation approaches are proposed. We take empirical experiments to analyze their impacts on the effects and efficiency of semantic matching respectively. The results are displayed in Table~\ref{tab:stopping_conditions}. For all cases, the max number of selected neighbors is restricted to 20.

First of all, setting a relatively small fixed capacity as the truncation shows very stable performance on various matching models. For the documents with few informative neighbors, these neighbors are involved without too much noise. For those documents with a lot of effective neighbors, selecting a part of top-tanked neighbors may be sufficient to distinguish them from the negative keys. Introducing adaptive strategies further reduces the computation cost, even promoting the matching accuracy. By setting an appropriate absolute value threshold, we could adaptively select informative neighbors for each pair of matching documents. Fewer neighbors will be selected for the scenarios where only a small fraction of neighbors are informative, instead of a fixed number, thus saving a lot of computing resources. More neighbors would be selected for the documents that have many informative neighbors, providing more supporting evidence to enhance the matching. The relevance score threshold shows the best performance in most cases. Recall that in Definition~\ref{def:1}, we argue that a neighbor contributes to a specific document matching task when it provides additional evidence. As such, the relevance score threshold performs as a serious filter to highlight neighbors that are more informative than the center document.

\begin{table}[t]
    \centering
    \caption{Comparison about the time and space cost of the lightweight selector and high-capacity semantic matching network.}
    \label{tab:cost_comparison}
    \begin{tabular}{p{0.2\linewidth}p{0.15\linewidth}p{0.16\linewidth}p{0.15\linewidth}p{0.16\linewidth}}
        \toprule
        \multirow{2}*{Scenarios} & \multicolumn{2}{c}{All 50 Neighbors} & \multicolumn{2}{c}{Select 5 Neighbors} \\
        \cmidrule(lr){2-3}
        \cmidrule(lr){4-5}
        & Time(ms) & Space(MB) & Time(ms) & Space(MB) \\
        \hline
        Selection & - & - & 1.72 & 1000  \\
        Matching & 1425.0 & 10219 & 217.0 & 2361 \\
        \hline
        Total & 1425.0 & 10219 & 218.72 & 3361 \\
        \bottomrule
    \end{tabular}
\end{table}

\begin{figure}
    \centering
    \includegraphics[width=0.7\linewidth]{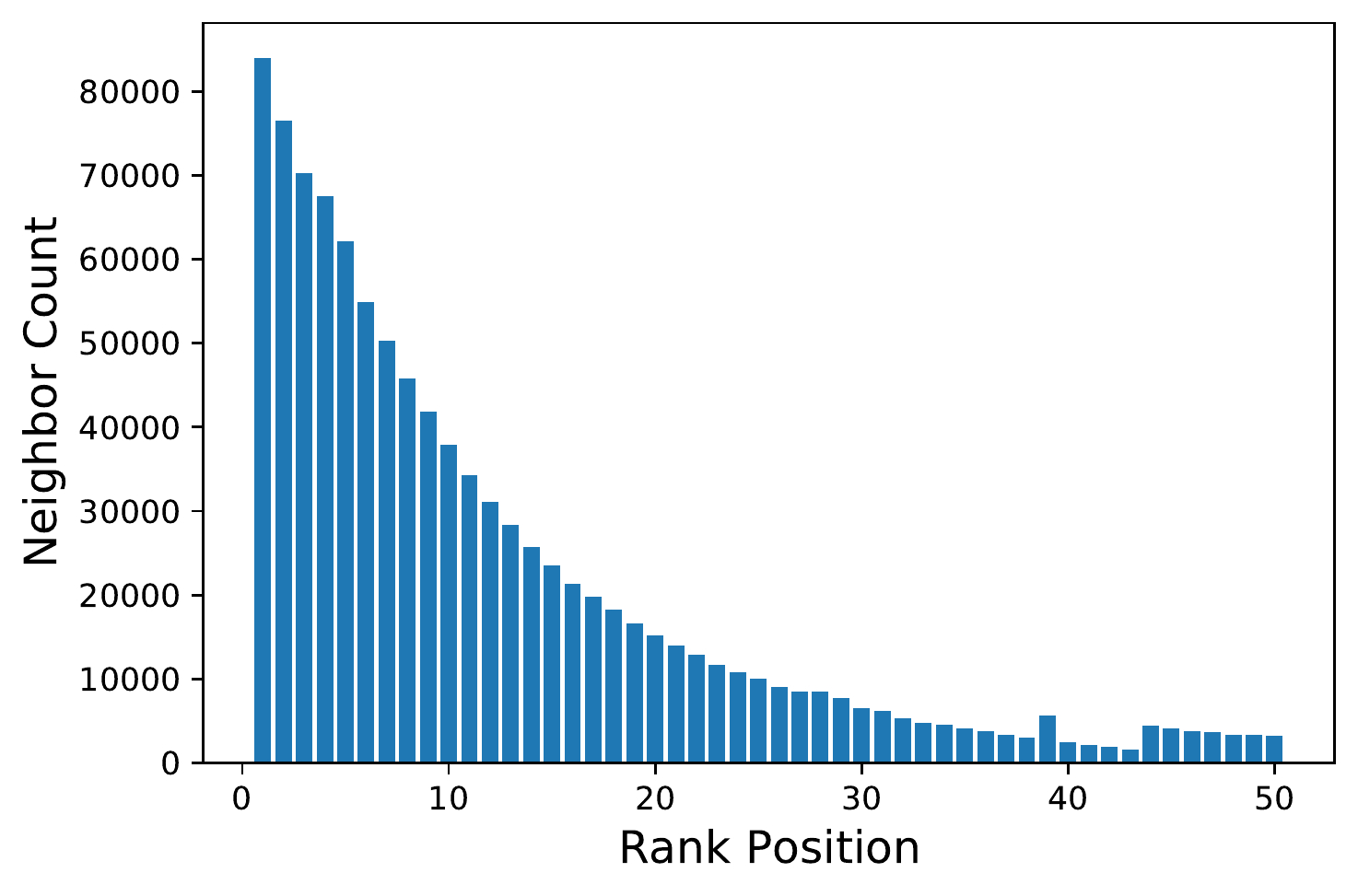}
    \caption{Distribution of the top 10 neighbors selected in the CDSM framework. The rank position is decided by the usefulness annotated by the well-trained semantic matching model.}
    \label{fig:selected_neighbors}
\end{figure}

\subsection{Effects and Efficiency of Selector (RQ3)}~\label{subsec:rq3}
To ensure the running efficiency, the neighbor selector in our CDSM framework should be lightweight. We compare its computation cost with the high-capacity matching network and present the results in Table~\ref{tab:cost_comparison}. The time and space cost of the selector are both much smaller than the expensive matching model, which could be ignored. Then, we conduct an experiment to analyze whether the highly simplified selector can learn to accurately identify useful neighbors.

Due to a lack of labels to explicitly measure the usefulness of neighbors, we develop a weak-supervision strategy to train the neighbor selector on top of weak annotations generated by the well-trained matching model. Thus, we evaluate the learning ability of the simplified selector by how well it could adapt to the annotations from the matching model. We make a comparative analysis of their respective evaluations about the neighbors. At first, we rank all neighbors of each matching task according to their usefulness measured by the semantic matching model as Definition~\ref{def:1}. Then, we apply the trained selector to score the neighbors and observe which neighbors in the above ranking are scored as the top 10. The distribution of original ranking positions is displayed in Figure~\ref{fig:selected_neighbors}. We find that most of the top 10 neighbors scored by the selector still correspond to those ranked at the top 10 positions by the high-capacity matching model, some at 10-20. This demonstrates that the highly simplified selector fits well to the annotations provided by the high-capacity matching model. It can identify useful neighbors efficiently and accurately.

\begin{figure}
    \centering
    \includegraphics[width=0.8\linewidth]{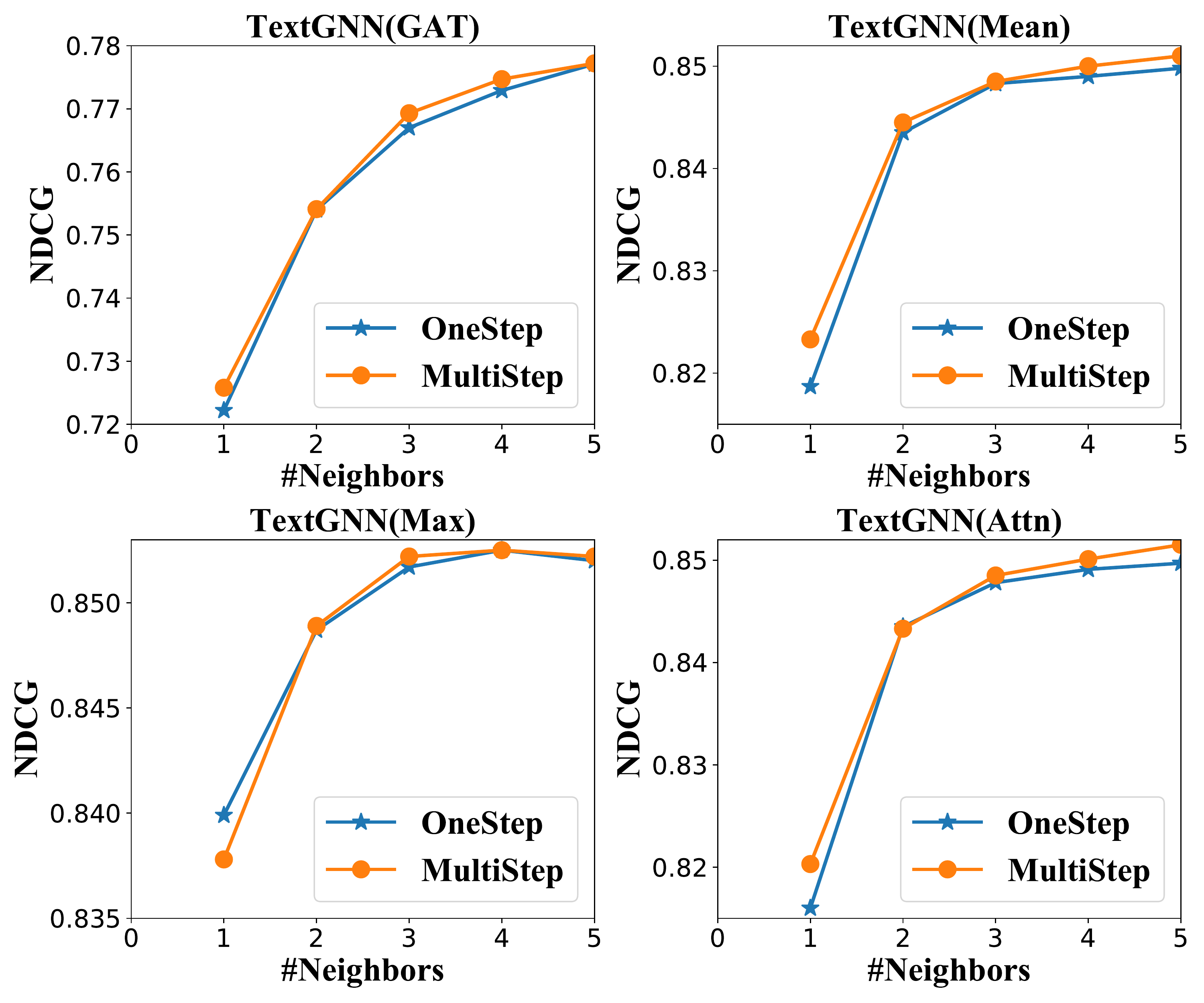}
    \caption{Performance curves of One-Step Selector and Multi-Step Selector on the Wiki dataset.}
    \label{fig:onestep_multistep}
\end{figure}

\subsection{One-Step v.s. Multi-Step Selection (RQ4)}~\label{subsec:rq4}
As for the ad-hoc neighbor selector in CDSM, we propose two different ranking functions, i.e. one-step and multi-step ranking functions. The one-step ranking function measures the usefulness of all neighbors based on only $Q, K$ and selects the top-$k$ highest-scoring neighbors at one step. The multi-step ranking function ranks and selects neighbors one by one. In each step, the neighbor leading to the largest information gain is selected. We experiment to compare these two selectors with TextGNN. Limited by the computing resources, at most 5 neighbors are selected.

Observing the performance curve of the two ranking functions in Figure~\ref{fig:onestep_multistep}, both approaches have the ability to distinguish the informativeness of different neighbors and select effective neighbors for the matching task. The multi-step ranking function performs a little better than the one-step. We infer that it may be because the multi-step method selects neighbors in a greedy way to construct a best-performed neighbor subset, instead of evaluating each neighbor separately. In this case, the computation cost of multi-step selection would be much larger than the one-step selection since it needs to evaluate the neighbors multiple times. On the whole, the more efficient one-step selection can achieve a better trade-off between effects and efficiency. We employ the one-step ranking function in other studies.

\subsection{Case Study}~\label{subsec:case_study}
\begin{figure}
    \centering
    \includegraphics[width=0.99\textwidth]{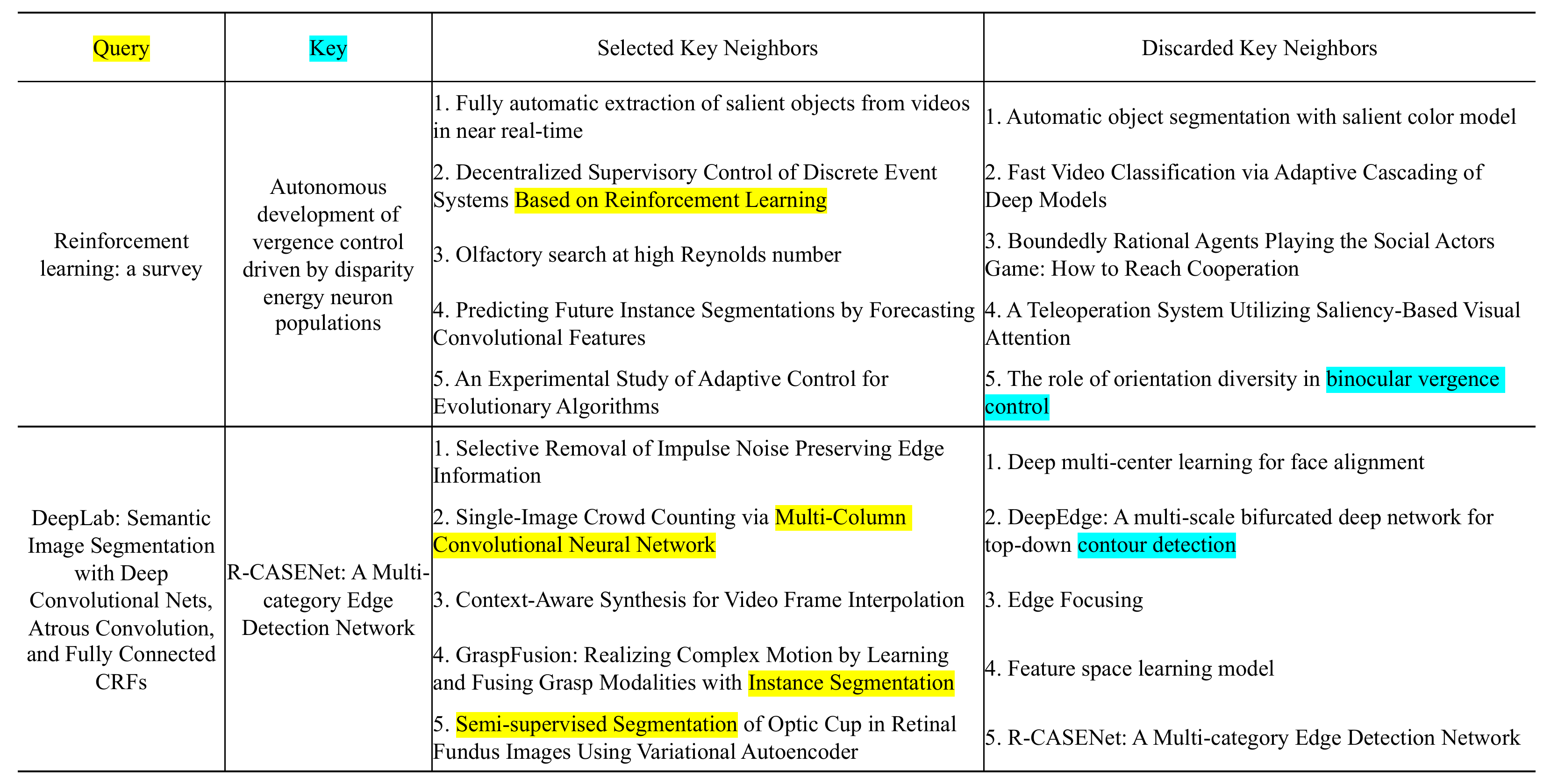}
    \caption{Case study of the selected and discarded neighbor nodes of the key document for the corresponding query. ``yellow'' highlights relevance with the query and ``blue'' highlights relevance with the key.}
    \label{fig:case_study}
\end{figure}

The neighbor selector is the core module in our proposed CDSM framework, and the quality of selected neighbors would directly affect the accuracy of the subsequent text matching. In order to evaluate the quality of neighbor selection, we carry out a qualitative case study on the DBLP dataset. For each pair of documents (Query $Q$, Key $K$) to be matched, we adopt the neighbor selector to score all neighbor nodes. Then, we compare the information contributed to the current matching task by the top 5 neighbors and the worst 5 neighbors respectively. Experimental results are presented in Figure~\ref{fig:case_study}.

We make the following observations. As for the neighbors highly scored by the selector, they can contribute some valuable information for determining the relationship between the current query and key documents. For example, facing the query ``Reinforcement learning: a survey'', the neighbor selector highlights the node relevant with ``reinforcement learning'' for discrimination. As for the query ``'DeepLab: Semantic Image Segmentation...', neighbors related to image segmentation are selected. However, the low-scoring neighbor nodes are much less relevant to the query but only relevant to the key, and even introduce noisy signals.

\section{Conclusion} \label{sec:conclusion}
In this paper, we identify the importance of neighbor selection in the graph-based semantic matching task and propose a novel framework, Cascaded Deep Semantic Matching (CDSM). It leverages a two-stage cascaded workflow for the semantic matching task. A lightweight neighbor selector is employed to efficiently filter informative neighbors for the given matching documents in the first stage. Then, the matching model calculates fine-grained relevance scores based on these selected neighbors. Both higher matching accuracy and speed are achieved by our CDSM framework, as unnecessary computation and background noise from irrelevant neighbors are avoided. We develop CDSM as a generic framework that can be seamlessly incorporated with mainstream graph-based models. A weak-supervision strategy is proposed to train the selector with weak annotations generated by the matching model. Empirical experiments on three large-scale textual graph datasets confirm the effectiveness and generality of our CDSM framework.

In summary, one limitation of the current CDSM is that the neighbor selector can accurately compare the usefulness of neighbors to highlight informative ones, but presents weak performance on truncation, i.e. determining how many neighbors should be selected. In Section~\ref{subsec:cdsm_framework}, we discuss several truncation approaches, without one consistently performing the best on all datasets. Designing a more universal automatic truncation method is a research topic in the future. In addition, the current two-step optimization method of CDSM can be iterated multiple times. How to enhance the ability of both the selector and matching model through iterative optimization is also under-explored.

\section*{Acknowledgment}

This work was supported by the National Natural Science Foundation of China No. 62272467 and No. U2001212, Beijing Outstanding Young Scientist Program NO. BJJWZYJH012019100020098, Public Computing Cloud, Renmin University of China. The work was partially done at Beijing Key Laboratory of Big Data Management and Analysis Methods, and Key Laboratory of Data Engineering and Knowledge Engineering, MOE.

% This work was supported by National Natural Science Foundation of China No. 61872370 and 61832017, Beijing Outstanding Young Scientist Program No. BJJWZYJH012019100020098 and in part by Shandong Provincial Natural Science Foundation under Grant ZR2019ZD06.

\newpage
\bibliographystyle{ACM-Reference-Format}
\bibliography{main}


\begin{thebibliography}{32}

%%% the Web address.

\ifx \showCODEN    \undefined \def \showCODEN     #1{\unskip}     \fi
\ifx \showDOI      \undefined \def \showDOI       #1{#1}\fi
\ifx \showISBNx    \undefined \def \showISBNx     #1{\unskip}     \fi
\ifx \showISBNxiii \undefined \def \showISBNxiii  #1{\unskip}     \fi
\ifx \showISSN     \undefined \def \showISSN      #1{\unskip}     \fi
\ifx \showLCCN     \undefined \def \showLCCN      #1{\unskip}     \fi
\ifx \shownote     \undefined \def \shownote      #1{#1}          \fi
\ifx \showarticletitle \undefined \def \showarticletitle #1{#1}   \fi
\ifx \showURL      \undefined \def \showURL       {\relax}        \fi
% invisible to TeX
\providecommand\bibfield[2]{#2}
\providecommand\bibinfo[2]{#2}
\providecommand\natexlab[1]{#1}
\providecommand\showeprint[2][]{arXiv:#2}

\bibitem[\protect\citeauthoryear{Arora}{Arora}{2020}]%
        {Arora2020KGcompletionSurvey}
\bibfield{author}{\bibinfo{person}{Siddhant Arora}.}
  \bibinfo{year}{2020}\natexlab{}.
\newblock \showarticletitle{A Survey on Graph Neural Networks for Knowledge
  Graph Completion}.
\newblock \bibinfo{journal}{\emph{CoRR}}  \bibinfo{volume}{abs/2007.12374}
  (\bibinfo{year}{2020}).
\newblock


\bibitem[\protect\citeauthoryear{Bao, Dong, Wei, Wang, Yang, Liu, Wang, Gao,
  Piao, Zhou, and Hon}{Bao et~al\mbox{.}}{2020}]%
        {Bao2020UniLMv2}
\bibfield{author}{\bibinfo{person}{Hangbo Bao}, \bibinfo{person}{Li Dong},
  \bibinfo{person}{Furu Wei}, \bibinfo{person}{Wenhui Wang},
  \bibinfo{person}{Nan Yang}, \bibinfo{person}{Xiaodong Liu},
  \bibinfo{person}{Yu Wang}, \bibinfo{person}{Jianfeng Gao},
  \bibinfo{person}{Songhao Piao}, \bibinfo{person}{Ming Zhou}, {and}
  \bibinfo{person}{Hsiao{-}Wuen Hon}.} \bibinfo{year}{2020}\natexlab{}.
\newblock \showarticletitle{UniLMv2: Pseudo-Masked Language Models for Unified
  Language Model Pre-Training}. In \bibinfo{booktitle}{\emph{Proceedings of the
  37th International Conference on Machine Learning, {ICML} 2020, 13-18 July
  2020, Virtual Event}} \emph{(\bibinfo{series}{Proceedings of Machine Learning
  Research}, Vol.~\bibinfo{volume}{119})}. \bibinfo{publisher}{{PMLR}},
  \bibinfo{pages}{642--652}.
\newblock


\bibitem[\protect\citeauthoryear{Blei, Ng, and Jordan}{Blei
  et~al\mbox{.}}{2003}]%
        {blei2003latent}
\bibfield{author}{\bibinfo{person}{David~M Blei}, \bibinfo{person}{Andrew~Y
  Ng}, {and} \bibinfo{person}{Michael~I Jordan}.}
  \bibinfo{year}{2003}\natexlab{}.
\newblock \showarticletitle{Latent dirichlet allocation}.
\newblock \bibinfo{journal}{\emph{the Journal of machine Learning research}}
  \bibinfo{volume}{3} (\bibinfo{year}{2003}), \bibinfo{pages}{993--1022}.
\newblock


\bibitem[\protect\citeauthoryear{Devlin, Chang, Lee, and Toutanova}{Devlin
  et~al\mbox{.}}{2019}]%
        {Devlin2019BERT}
\bibfield{author}{\bibinfo{person}{Jacob Devlin}, \bibinfo{person}{Ming{-}Wei
  Chang}, \bibinfo{person}{Kenton Lee}, {and} \bibinfo{person}{Kristina
  Toutanova}.} \bibinfo{year}{2019}\natexlab{}.
\newblock \showarticletitle{{BERT:} Pre-training of Deep Bidirectional
  Transformers for Language Understanding}. In
  \bibinfo{booktitle}{\emph{Proceedings of the 2019 Conference of the North
  American Chapter of the Association for Computational Linguistics: Human
  Language Technologies, {NAACL-HLT} 2019, Minneapolis, MN, USA, June 2-7,
  2019, Volume 1 (Long and Short Papers)}}. \bibinfo{publisher}{Association for
  Computational Linguistics}, \bibinfo{pages}{4171--4186}.
\newblock


\bibitem[\protect\citeauthoryear{Guo, Fan, Ai, and Croft}{Guo
  et~al\mbox{.}}{2016}]%
        {guo2016deep}
\bibfield{author}{\bibinfo{person}{Jiafeng Guo}, \bibinfo{person}{Yixing Fan},
  \bibinfo{person}{Qingyao Ai}, {and} \bibinfo{person}{W~Bruce Croft}.}
  \bibinfo{year}{2016}\natexlab{}.
\newblock \showarticletitle{A deep relevance matching model for ad-hoc
  retrieval}. In \bibinfo{booktitle}{\emph{Proceedings of the 25th ACM
  international on conference on information and knowledge management}}.
  \bibinfo{pages}{55--64}.
\newblock


\bibitem[\protect\citeauthoryear{Hamilton, Ying, and Leskovec}{Hamilton
  et~al\mbox{.}}{2017a}]%
        {hamilton2017inductive}
\bibfield{author}{\bibinfo{person}{William~L Hamilton}, \bibinfo{person}{Rex
  Ying}, {and} \bibinfo{person}{Jure Leskovec}.}
  \bibinfo{year}{2017}\natexlab{a}.
\newblock \showarticletitle{Inductive representation learning on large graphs}.
  In \bibinfo{booktitle}{\emph{Proceedings of the 31st International Conference
  on Neural Information Processing Systems}}. \bibinfo{pages}{1025--1035}.
\newblock


\bibitem[\protect\citeauthoryear{Hamilton, Ying, and Leskovec}{Hamilton
  et~al\mbox{.}}{2017b}]%
        {Hamilton2017GraphSage}
\bibfield{author}{\bibinfo{person}{William~L. Hamilton},
  \bibinfo{person}{Zhitao Ying}, {and} \bibinfo{person}{Jure Leskovec}.}
  \bibinfo{year}{2017}\natexlab{b}.
\newblock \showarticletitle{Inductive Representation Learning on Large Graphs}.
  In \bibinfo{booktitle}{\emph{Advances in Neural Information Processing
  Systems 30: Annual Conference on Neural Information Processing Systems 2017,
  December 4-9, 2017, Long Beach, CA, {USA}}}. \bibinfo{pages}{1024--1034}.
\newblock


\bibitem[\protect\citeauthoryear{Hu, Xu, Li, Yang, Shi, Duan, Xie, and Zhou}{Hu
  et~al\mbox{.}}{2020}]%
        {Hu2020RecoGNN}
\bibfield{author}{\bibinfo{person}{Linmei Hu}, \bibinfo{person}{Siyong Xu},
  \bibinfo{person}{Chen Li}, \bibinfo{person}{Cheng Yang},
  \bibinfo{person}{Chuan Shi}, \bibinfo{person}{Nan Duan},
  \bibinfo{person}{Xing Xie}, {and} \bibinfo{person}{Ming Zhou}.}
  \bibinfo{year}{2020}\natexlab{}.
\newblock \showarticletitle{Graph Neural News Recommendation with Unsupervised
  Preference Disentanglement}. In \bibinfo{booktitle}{\emph{Proceedings of the
  58th Annual Meeting of the Association for Computational Linguistics, {ACL}
  2020, Online, July 5-10, 2020}}. \bibinfo{publisher}{Association for
  Computational Linguistics}, \bibinfo{pages}{4255--4264}.
\newblock


\bibitem[\protect\citeauthoryear{Huang, He, Gao, Deng, Acero, and Heck}{Huang
  et~al\mbox{.}}{2013}]%
        {huang2013learning}
\bibfield{author}{\bibinfo{person}{Po-Sen Huang}, \bibinfo{person}{Xiaodong
  He}, \bibinfo{person}{Jianfeng Gao}, \bibinfo{person}{Li Deng},
  \bibinfo{person}{Alex Acero}, {and} \bibinfo{person}{Larry Heck}.}
  \bibinfo{year}{2013}\natexlab{}.
\newblock \showarticletitle{Learning deep structured semantic models for web
  search using clickthrough data}. In \bibinfo{booktitle}{\emph{Proceedings of
  the 22nd ACM international conference on Information \& Knowledge
  Management}}. \bibinfo{pages}{2333--2338}.
\newblock


\bibitem[\protect\citeauthoryear{Karpukhin, Oguz, Min, Lewis, Wu, Edunov, Chen,
  and Yih}{Karpukhin et~al\mbox{.}}{2020}]%
        {Karpukhin2020inbatch}
\bibfield{author}{\bibinfo{person}{Vladimir Karpukhin}, \bibinfo{person}{Barlas
  Oguz}, \bibinfo{person}{Sewon Min}, \bibinfo{person}{Patrick S.~H. Lewis},
  \bibinfo{person}{Ledell Wu}, \bibinfo{person}{Sergey Edunov},
  \bibinfo{person}{Danqi Chen}, {and} \bibinfo{person}{Wen{-}tau Yih}.}
  \bibinfo{year}{2020}\natexlab{}.
\newblock \showarticletitle{Dense Passage Retrieval for Open-Domain Question
  Answering}. In \bibinfo{booktitle}{\emph{Proceedings of the 2020 Conference
  on Empirical Methods in Natural Language Processing, {EMNLP} 2020, Online,
  November 16-20, 2020}}. \bibinfo{pages}{6769--6781}.
\newblock


\bibitem[\protect\citeauthoryear{Khattab and Zaharia}{Khattab and
  Zaharia}{2020}]%
        {khattab2020colbert}
\bibfield{author}{\bibinfo{person}{Omar Khattab} {and} \bibinfo{person}{Matei
  Zaharia}.} \bibinfo{year}{2020}\natexlab{}.
\newblock \showarticletitle{Colbert: Efficient and effective passage search via
  contextualized late interaction over bert}. In
  \bibinfo{booktitle}{\emph{Proceedings of the 43rd International ACM SIGIR
  Conference on Research and Development in Information Retrieval}}.
  \bibinfo{pages}{39--48}.
\newblock


\bibitem[\protect\citeauthoryear{Landauer, Foltz, and Laham}{Landauer
  et~al\mbox{.}}{1998}]%
        {landauer1998introduction}
\bibfield{author}{\bibinfo{person}{Thomas~K Landauer}, \bibinfo{person}{Peter~W
  Foltz}, {and} \bibinfo{person}{Darrell Laham}.}
  \bibinfo{year}{1998}\natexlab{}.
\newblock \showarticletitle{An introduction to latent semantic analysis}.
\newblock \bibinfo{journal}{\emph{Discourse processes}} \bibinfo{volume}{25},
  \bibinfo{number}{2-3} (\bibinfo{year}{1998}), \bibinfo{pages}{259--284}.
\newblock


\bibitem[\protect\citeauthoryear{Li, Pang, Liu, Sun, Liu, Xie, Yang, Cui,
  Zhang, and Zhang}{Li et~al\mbox{.}}{2021}]%
        {Li2021AdsGNN}
\bibfield{author}{\bibinfo{person}{Chaozhuo Li}, \bibinfo{person}{Bochen Pang},
  \bibinfo{person}{Yuming Liu}, \bibinfo{person}{Hao Sun},
  \bibinfo{person}{Zheng Liu}, \bibinfo{person}{Xing Xie},
  \bibinfo{person}{Tianqi Yang}, \bibinfo{person}{Yanling Cui},
  \bibinfo{person}{Liangjie Zhang}, {and} \bibinfo{person}{Qi Zhang}.}
  \bibinfo{year}{2021}\natexlab{}.
\newblock \showarticletitle{AdsGNN: Behavior-Graph Augmented Relevance Modeling
  in Sponsored Search}. In \bibinfo{booktitle}{\emph{{SIGIR} '21: The 44th
  International {ACM} {SIGIR} Conference on Research and Development in
  Information Retrieval, Virtual Event, Canada, July 11-15, 2021}},
  \bibfield{editor}{\bibinfo{person}{Fernando Diaz}, \bibinfo{person}{Chirag
  Shah}, \bibinfo{person}{Torsten Suel}, \bibinfo{person}{Pablo Castells},
  \bibinfo{person}{Rosie Jones}, {and} \bibinfo{person}{Tetsuya Sakai}} (Eds.).
  \bibinfo{publisher}{{ACM}}, \bibinfo{pages}{223--232}.
\newblock


\bibitem[\protect\citeauthoryear{Liu, Ott, Goyal, Du, Joshi, Chen, Levy, Lewis,
  Zettlemoyer, and Stoyanov}{Liu et~al\mbox{.}}{2019}]%
        {Liu2019Roberta}
\bibfield{author}{\bibinfo{person}{Yinhan Liu}, \bibinfo{person}{Myle Ott},
  \bibinfo{person}{Naman Goyal}, \bibinfo{person}{Jingfei Du},
  \bibinfo{person}{Mandar Joshi}, \bibinfo{person}{Danqi Chen},
  \bibinfo{person}{Omer Levy}, \bibinfo{person}{Mike Lewis},
  \bibinfo{person}{Luke Zettlemoyer}, {and} \bibinfo{person}{Veselin
  Stoyanov}.} \bibinfo{year}{2019}\natexlab{}.
\newblock \showarticletitle{RoBERTa: {A} Robustly Optimized {BERT} Pretraining
  Approach}.
\newblock \bibinfo{journal}{\emph{CoRR}}  \bibinfo{volume}{abs/1907.11692}
  (\bibinfo{year}{2019}).
\newblock


\bibitem[\protect\citeauthoryear{Luan, Eisenstein, Toutanova, and Collins}{Luan
  et~al\mbox{.}}{2021}]%
        {luan2020sparse}
\bibfield{author}{\bibinfo{person}{Yi Luan}, \bibinfo{person}{Jacob
  Eisenstein}, \bibinfo{person}{Kristina Toutanova}, {and}
  \bibinfo{person}{Michael Collins}.} \bibinfo{year}{2021}\natexlab{}.
\newblock \showarticletitle{Sparse, Dense, and Attentional Representations for
  Text Retrieval}.
\newblock \bibinfo{journal}{\emph{Trans. Assoc. Comput. Linguistics}}
  \bibinfo{volume}{9} (\bibinfo{year}{2021}), \bibinfo{pages}{329--345}.
\newblock


\bibitem[\protect\citeauthoryear{Palangi, Deng, Shen, Gao, He, Chen, Song, and
  Ward}{Palangi et~al\mbox{.}}{2016}]%
        {palangi2016deep}
\bibfield{author}{\bibinfo{person}{Hamid Palangi}, \bibinfo{person}{Li Deng},
  \bibinfo{person}{Yelong Shen}, \bibinfo{person}{Jianfeng Gao},
  \bibinfo{person}{Xiaodong He}, \bibinfo{person}{Jianshu Chen},
  \bibinfo{person}{Xinying Song}, {and} \bibinfo{person}{Rabab Ward}.}
  \bibinfo{year}{2016}\natexlab{}.
\newblock \showarticletitle{Deep sentence embedding using long short-term
  memory networks: Analysis and application to information retrieval}.
\newblock \bibinfo{journal}{\emph{IEEE/ACM Transactions on Audio, Speech, and
  Language Processing}} \bibinfo{volume}{24}, \bibinfo{number}{4}
  (\bibinfo{year}{2016}), \bibinfo{pages}{694--707}.
\newblock


\bibitem[\protect\citeauthoryear{Reimers and Gurevych}{Reimers and
  Gurevych}{2019}]%
        {reimers2019sentence}
\bibfield{author}{\bibinfo{person}{Nils Reimers} {and} \bibinfo{person}{Iryna
  Gurevych}.} \bibinfo{year}{2019}\natexlab{}.
\newblock \showarticletitle{Sentence-BERT: Sentence Embeddings using Siamese
  BERT-Networks}. In \bibinfo{booktitle}{\emph{Proceedings of the 2019
  Conference on Empirical Methods in Natural Language Processing and the 9th
  International Joint Conference on Natural Language Processing, {EMNLP-IJCNLP}
  2019, Hong Kong, China, November 3-7, 2019}},
  \bibfield{editor}{\bibinfo{person}{Kentaro Inui}, \bibinfo{person}{Jing
  Jiang}, \bibinfo{person}{Vincent Ng}, {and} \bibinfo{person}{Xiaojun Wan}}
  (Eds.). \bibinfo{publisher}{Association for Computational Linguistics},
  \bibinfo{pages}{3980--3990}.
\newblock


\bibitem[\protect\citeauthoryear{Robertson and Zaragoza}{Robertson and
  Zaragoza}{2009}]%
        {robertson2009probabilistic}
\bibfield{author}{\bibinfo{person}{Stephen Robertson} {and}
  \bibinfo{person}{Hugo Zaragoza}.} \bibinfo{year}{2009}\natexlab{}.
\newblock \bibinfo{booktitle}{\emph{The probabilistic relevance framework: BM25
  and beyond}}.
\newblock \bibinfo{publisher}{Now Publishers Inc}.
\newblock


\bibitem[\protect\citeauthoryear{Seo, Kembhavi, Farhadi, and Hajishirzi}{Seo
  et~al\mbox{.}}{2017}]%
        {Seo2017BiDAF}
\bibfield{author}{\bibinfo{person}{Min~Joon Seo}, \bibinfo{person}{Aniruddha
  Kembhavi}, \bibinfo{person}{Ali Farhadi}, {and} \bibinfo{person}{Hannaneh
  Hajishirzi}.} \bibinfo{year}{2017}\natexlab{}.
\newblock \showarticletitle{Bidirectional Attention Flow for Machine
  Comprehension}. In \bibinfo{booktitle}{\emph{5th International Conference on
  Learning Representations, {ICLR} 2017, Toulon, France, April 24-26, 2017,
  Conference Track Proceedings}}. \bibinfo{publisher}{OpenReview.net}.
\newblock


\bibitem[\protect\citeauthoryear{Shen, He, Gao, Deng, and Mesnil}{Shen
  et~al\mbox{.}}{2014}]%
        {shen2014learning}
\bibfield{author}{\bibinfo{person}{Yelong Shen}, \bibinfo{person}{Xiaodong He},
  \bibinfo{person}{Jianfeng Gao}, \bibinfo{person}{Li Deng}, {and}
  \bibinfo{person}{Gr{\'e}goire Mesnil}.} \bibinfo{year}{2014}\natexlab{}.
\newblock \showarticletitle{Learning semantic representations using
  convolutional neural networks for web search}. In
  \bibinfo{booktitle}{\emph{Proceedings of the 23rd international conference on
  world wide web}}. \bibinfo{pages}{373--374}.
\newblock


\bibitem[\protect\citeauthoryear{Velickovic, Cucurull, Casanova, Romero,
  Li{\`{o}}, and Bengio}{Velickovic et~al\mbox{.}}{2018}]%
        {Velickovic2017GAT}
\bibfield{author}{\bibinfo{person}{Petar Velickovic}, \bibinfo{person}{Guillem
  Cucurull}, \bibinfo{person}{Arantxa Casanova}, \bibinfo{person}{Adriana
  Romero}, \bibinfo{person}{Pietro Li{\`{o}}}, {and} \bibinfo{person}{Yoshua
  Bengio}.} \bibinfo{year}{2018}\natexlab{}.
\newblock \showarticletitle{Graph Attention Networks}. In
  \bibinfo{booktitle}{\emph{6th International Conference on Learning
  Representations, {ICLR} 2018, Vancouver, BC, Canada, April 30 - May 3, 2018,
  Conference Track Proceedings}}. \bibinfo{publisher}{OpenReview.net}.
\newblock


\bibitem[\protect\citeauthoryear{Wang, Huang, Zhao, Zhang, Zhao, and Lee}{Wang
  et~al\mbox{.}}{2018}]%
        {wang2018billion}
\bibfield{author}{\bibinfo{person}{Jizhe Wang}, \bibinfo{person}{Pipei Huang},
  \bibinfo{person}{Huan Zhao}, \bibinfo{person}{Zhibo Zhang},
  \bibinfo{person}{Binqiang Zhao}, {and} \bibinfo{person}{Dik~Lun Lee}.}
  \bibinfo{year}{2018}\natexlab{}.
\newblock \showarticletitle{Billion-scale commodity embedding for e-commerce
  recommendation in alibaba}. In \bibinfo{booktitle}{\emph{Proceedings of the
  24th ACM SIGKDD International Conference on Knowledge Discovery \& Data
  Mining}}. \bibinfo{pages}{839--848}.
\newblock


\bibitem[\protect\citeauthoryear{Wang, Qiu, and Wang}{Wang
  et~al\mbox{.}}{2021b}]%
        {Wang2021KGCompletionSurvey}
\bibfield{author}{\bibinfo{person}{Meihong Wang}, \bibinfo{person}{Linling
  Qiu}, {and} \bibinfo{person}{Xiaoli Wang}.} \bibinfo{year}{2021}\natexlab{b}.
\newblock \showarticletitle{A Survey on Knowledge Graph Embeddings for Link
  Prediction}.
\newblock \bibinfo{journal}{\emph{Symmetry}} \bibinfo{volume}{13},
  \bibinfo{number}{3} (\bibinfo{year}{2021}), \bibinfo{pages}{485}.
\newblock


\bibitem[\protect\citeauthoryear{Wang, Gao, Zhu, Zhang, Liu, Li, and Tang}{Wang
  et~al\mbox{.}}{2021a}]%
        {Wang2021Wikidata}
\bibfield{author}{\bibinfo{person}{Xiaozhi Wang}, \bibinfo{person}{Tianyu Gao},
  \bibinfo{person}{Zhaocheng Zhu}, \bibinfo{person}{Zhengyan Zhang},
  \bibinfo{person}{Zhiyuan Liu}, \bibinfo{person}{Juanzi Li}, {and}
  \bibinfo{person}{Jian Tang}.} \bibinfo{year}{2021}\natexlab{a}.
\newblock \showarticletitle{{KEPLER:} {A} Unified Model for Knowledge Embedding
  and Pre-trained Language Representation}.
\newblock \bibinfo{journal}{\emph{Trans. Assoc. Comput. Linguistics}}
  \bibinfo{volume}{9} (\bibinfo{year}{2021}), \bibinfo{pages}{176--194}.
\newblock
\urldef\tempurl%
\url{https://transacl.org/ojs/index.php/tacl/article/view/2447}
\showURL{%
\tempurl}


\bibitem[\protect\citeauthoryear{Wu, Schuster, Chen, Le, Norouzi, Macherey,
  Krikun, Cao, Gao, Macherey, Klingner, Shah, Johnson, Liu, Kaiser, Gouws,
  Kato, Kudo, Kazawa, Stevens, Kurian, Patil, Wang, Young, Smith, Riesa,
  Rudnick, Vinyals, Corrado, Hughes, and Dean}{Wu et~al\mbox{.}}{2016}]%
        {Wu2016google}
\bibfield{author}{\bibinfo{person}{Yonghui Wu}, \bibinfo{person}{Mike
  Schuster}, \bibinfo{person}{Zhifeng Chen}, \bibinfo{person}{Quoc~V. Le},
  \bibinfo{person}{Mohammad Norouzi}, \bibinfo{person}{Wolfgang Macherey},
  \bibinfo{person}{Maxim Krikun}, \bibinfo{person}{Yuan Cao},
  \bibinfo{person}{Qin Gao}, \bibinfo{person}{Klaus Macherey},
  \bibinfo{person}{Jeff Klingner}, \bibinfo{person}{Apurva Shah},
  \bibinfo{person}{Melvin Johnson}, \bibinfo{person}{Xiaobing Liu},
  \bibinfo{person}{Lukasz Kaiser}, \bibinfo{person}{Stephan Gouws},
  \bibinfo{person}{Yoshikiyo Kato}, \bibinfo{person}{Taku Kudo},
  \bibinfo{person}{Hideto Kazawa}, \bibinfo{person}{Keith Stevens},
  \bibinfo{person}{George Kurian}, \bibinfo{person}{Nishant Patil},
  \bibinfo{person}{Wei Wang}, \bibinfo{person}{Cliff Young},
  \bibinfo{person}{Jason Smith}, \bibinfo{person}{Jason Riesa},
  \bibinfo{person}{Alex Rudnick}, \bibinfo{person}{Oriol Vinyals},
  \bibinfo{person}{Greg Corrado}, \bibinfo{person}{Macduff Hughes}, {and}
  \bibinfo{person}{Jeffrey Dean}.} \bibinfo{year}{2016}\natexlab{}.
\newblock \showarticletitle{Google's Neural Machine Translation System:
  Bridging the Gap between Human and Machine Translation}.
\newblock \bibinfo{journal}{\emph{CoRR}}  \bibinfo{volume}{abs/1609.08144}
  (\bibinfo{year}{2016}).
\newblock


\bibitem[\protect\citeauthoryear{Xu, He, and Li}{Xu et~al\mbox{.}}{2018}]%
        {xu2018deep}
\bibfield{author}{\bibinfo{person}{Jun Xu}, \bibinfo{person}{Xiangnan He},
  {and} \bibinfo{person}{Hang Li}.} \bibinfo{year}{2018}\natexlab{}.
\newblock \showarticletitle{Deep learning for matching in search and
  recommendation}. In \bibinfo{booktitle}{\emph{The 41st International ACM
  SIGIR Conference on Research \& Development in Information Retrieval}}.
  \bibinfo{pages}{1365--1368}.
\newblock


\bibitem[\protect\citeauthoryear{Yang, Liu, Xiao, Li, Lian, Agrawal, Amit, Sun,
  and Xie}{Yang et~al\mbox{.}}{2021}]%
        {Yang2021GraphFormer}
\bibfield{author}{\bibinfo{person}{Junhan Yang}, \bibinfo{person}{Zheng Liu},
  \bibinfo{person}{Shitao Xiao}, \bibinfo{person}{Chaozhuo Li},
  \bibinfo{person}{Defu Lian}, \bibinfo{person}{Sanjay Agrawal},
  \bibinfo{person}{S Amit}, \bibinfo{person}{Guangzhong Sun}, {and}
  \bibinfo{person}{Xing Xie}.} \bibinfo{year}{2021}\natexlab{}.
\newblock \showarticletitle{GraphFormers: GNN-nested Transformers for
  Representation Learning on Textual Graph}.
\newblock \bibinfo{journal}{\emph{Thirty-Fifth Conference on Neural Information
  Processing Systems}} (\bibinfo{year}{2021}).
\newblock


\bibitem[\protect\citeauthoryear{Yang, Yang, Dyer, He, Smola, and Hovy}{Yang
  et~al\mbox{.}}{2016}]%
        {yang2016hierarchical}
\bibfield{author}{\bibinfo{person}{Zichao Yang}, \bibinfo{person}{Diyi Yang},
  \bibinfo{person}{Chris Dyer}, \bibinfo{person}{Xiaodong He},
  \bibinfo{person}{Alex Smola}, {and} \bibinfo{person}{Eduard Hovy}.}
  \bibinfo{year}{2016}\natexlab{}.
\newblock \showarticletitle{Hierarchical attention networks for document
  classification}. In \bibinfo{booktitle}{\emph{Proceedings of the 2016
  conference of the North American chapter of the association for computational
  linguistics: human language technologies}}. \bibinfo{pages}{1480--1489}.
\newblock


\bibitem[\protect\citeauthoryear{Yao, Dou, and Wen}{Yao et~al\mbox{.}}{2020}]%
        {yao2020peps}
\bibfield{author}{\bibinfo{person}{Jing Yao}, \bibinfo{person}{Zhicheng Dou},
  {and} \bibinfo{person}{Ji{-}Rong Wen}.} \bibinfo{year}{2020}\natexlab{}.
\newblock \showarticletitle{Employing Personal Word Embeddings for Personalized
  Search}. In \bibinfo{booktitle}{\emph{Proceedings of the 43rd International
  {ACM} {SIGIR} conference on research and development in Information
  Retrieval, {SIGIR} 2020, Virtual Event, China, July 25-30, 2020}},
  \bibfield{editor}{\bibinfo{person}{Jimmy Huang}, \bibinfo{person}{Yi~Chang},
  \bibinfo{person}{Xueqi Cheng}, \bibinfo{person}{Jaap Kamps},
  \bibinfo{person}{Vanessa Murdock}, \bibinfo{person}{Ji{-}Rong Wen}, {and}
  \bibinfo{person}{Yiqun Liu}} (Eds.). \bibinfo{publisher}{{ACM}},
  \bibinfo{pages}{1359--1368}.
\newblock


\bibitem[\protect\citeauthoryear{Ying, He, Chen, Eksombatchai, Hamilton, and
  Leskovec}{Ying et~al\mbox{.}}{2018}]%
        {ying2018graph}
\bibfield{author}{\bibinfo{person}{Rex Ying}, \bibinfo{person}{Ruining He},
  \bibinfo{person}{Kaifeng Chen}, \bibinfo{person}{Pong Eksombatchai},
  \bibinfo{person}{William~L Hamilton}, {and} \bibinfo{person}{Jure Leskovec}.}
  \bibinfo{year}{2018}\natexlab{}.
\newblock \showarticletitle{Graph convolutional neural networks for web-scale
  recommender systems}. In \bibinfo{booktitle}{\emph{Proceedings of the 24th
  ACM SIGKDD International Conference on Knowledge Discovery \& Data Mining}}.
  \bibinfo{pages}{974--983}.
\newblock


\bibitem[\protect\citeauthoryear{Zhang, Liu, Luo, Xue, Shan, Luo, Xia, Yan, and
  Wang}{Zhang et~al\mbox{.}}{2021}]%
        {zhang2020mira}
\bibfield{author}{\bibinfo{person}{Yusi Zhang}, \bibinfo{person}{Chuanjie Liu},
  \bibinfo{person}{Angen Luo}, \bibinfo{person}{Hui Xue}, \bibinfo{person}{Xuan
  Shan}, \bibinfo{person}{Yuxiang Luo}, \bibinfo{person}{Yiqian Xia},
  \bibinfo{person}{Yuanchi Yan}, {and} \bibinfo{person}{Haidong Wang}.}
  \bibinfo{year}{2021}\natexlab{}.
\newblock \showarticletitle{{MIRA:} Leveraging Multi-Intention Co-click
  Information in Web-scale Document Retrieval using Deep Neural Networks}. In
  \bibinfo{booktitle}{\emph{{WWW} '21: The Web Conference 2021, Virtual Event /
  Ljubljana, Slovenia, April 19-23, 2021}},
  \bibfield{editor}{\bibinfo{person}{Jure Leskovec}, \bibinfo{person}{Marko
  Grobelnik}, \bibinfo{person}{Marc Najork}, \bibinfo{person}{Jie Tang}, {and}
  \bibinfo{person}{Leila Zia}} (Eds.). \bibinfo{publisher}{{ACM} / {IW3C2}},
  \bibinfo{pages}{227--238}.
\newblock


\bibitem[\protect\citeauthoryear{Zhu, Cui, Liu, Sun, Li, Pelger, Yang, Zhang,
  Zhang, and Zhao}{Zhu et~al\mbox{.}}{2021}]%
        {Zhu2021TextGNN}
\bibfield{author}{\bibinfo{person}{Jason Zhu}, \bibinfo{person}{Yanling Cui},
  \bibinfo{person}{Yuming Liu}, \bibinfo{person}{Hao Sun}, \bibinfo{person}{Xue
  Li}, \bibinfo{person}{Markus Pelger}, \bibinfo{person}{Tianqi Yang},
  \bibinfo{person}{Liangjie Zhang}, \bibinfo{person}{Ruofei Zhang}, {and}
  \bibinfo{person}{Huasha Zhao}.} \bibinfo{year}{2021}\natexlab{}.
\newblock \showarticletitle{TextGNN: Improving Text Encoder via Graph Neural
  Network in Sponsored Search}. In \bibinfo{booktitle}{\emph{{WWW} '21: The Web
  Conference 2021, Virtual Event / Ljubljana, Slovenia, April 19-23, 2021}},
  \bibfield{editor}{\bibinfo{person}{Jure Leskovec}, \bibinfo{person}{Marko
  Grobelnik}, \bibinfo{person}{Marc Najork}, \bibinfo{person}{Jie Tang}, {and}
  \bibinfo{person}{Leila Zia}} (Eds.). \bibinfo{publisher}{{ACM} / {IW3C2}},
  \bibinfo{pages}{2848--2857}.
\newblock


\end{thebibliography}

\end{document}